\documentclass{aa}  

\usepackage{graphicx}
\usepackage{booktabs}
\usepackage{subfigure} 
\usepackage{makecell}
\usepackage{ulem}
\usepackage{lineno}

\usepackage{txfonts}
\usepackage[colorlinks=true,linkcolor=blue,citecolor=blue]{hyperref}
\usepackage{xcolor}
\usepackage{orcidlink}
\definecolor{purple}{rgb}{0.5,0,0.87}

\newcommand{\samane}[1]{\textcolor{black}{#1}}

\begin{document}

   \title{Witnessing the rapid growth of disk galaxies over cosmic time using JWST and HST}

   \subtitle{}

   \author{Samane Raji\orcidlink{0000-0001-9000-5507}\inst{1,2}
        \and
          Mireia Montes\orcidlink{0000-0001-7847-0393}\inst{3}
        \and
          Ignacio Trujillo \orcidlink{0000-0001-8647-2874} \inst{4,5}
        \and
          Fernando Buitrago \orcidlink{0000-0002-2861-9812}\inst{1,2,6}
        \and 
          Carlos Marrero-de la Rosa \orcidlink{0009-0005-1728-8076} \inst{4,5}
        \and 
          Andrés Asensio Ramos \orcidlink{0000-0002-1248-0553} \inst{4,5}
        }

   \institute{Departamento de Física Teórica, Atómica y Óptica, Universidad de Valladolid, 47011, Valladolid, Spain
   \and Laboratory for Disruptive Interdisciplinary Science (LaDIS), Universidad de Valladolid, 47011 Valladolid, Spain
   \and Institute of Space Sciences (ICE, CSIC), Campus UAB, Carrer de Can Magrans, s/n, 08193 Barcelona, Spain   
   \and Instituto de Astrofísica de Canarias, c/Vía Láctea s/n, E38205 - La Laguna, Tenerife, Spain 
   \and Departamento de Astrofísica, Universidad de La Laguna, E-38205 - La Laguna, Tenerife, Spain
    \and Instituto de Astrofísica e Ciéncia do Espaço, Universidad de Lisboa, OAL, Tapada de Ajuda, PT1349-018 Lisbon, Portugal}

   \date{Received 13 November 2025; accepted 7 May 2026}

 
  \abstract
   {Measuring galaxy sizes is fundamental to understanding how galaxies grow and evolve. Traditional methods to measure sizes trace better the concentration of light (i.e., effective radius) or are limited by the depth of the survey (isophotal methods). With the advent of deep, wide surveys, a new and physically-motivated definition of size has arisen: the edge of the galaxy, which is the most distant location where star formation has or is occurring.
   In this work, we take advantage of the extraordinary depth and spatial resolution of the Hubble and James Webb Space Telescopes to do the most accurate study of galaxy edges at z$\sim$1. Using 22 photometric bands, we derived the radial age and metallicity profiles of two disk galaxies in the GOODS South field with stellar masses $\sim 4\times10^{10}$ $M_\odot$. The age profiles display a characteristic U-shape, while the metallicity profiles steadily decrease over galactocentric distance. The turnover of the age profile occurs near the galaxy’s edge, suggesting that stellar migration is responsible for the stars beyond the edge of these galaxies.
   Comparing to $z=0$ disk galaxies reveals that $z=1$  galaxies will grow inside-out, with little to no increase in mass in the inner 8 kpc, but a significant ($\sim 300\%$) increase in the outer parts.
   }



   \titlerunning{MW-like galaxies with JWST and HST}
   \authorrunning{Raji et al.}

   \maketitle
   \nolinenumbers

\section{Introduction}

The most widely used indicator of galaxy size is the effective radius (R$_e$), which is the radius enclosing half of a galaxy's total light. Studies using this size indicator \citep[e.g.,][]{shen2003, vanderwel2014, lange2015}, have provided the size evolution of galaxies of different masses and morphologies. For disk galaxies with a similar stellar mass to the Milky Way (MW), the size evolution based on $R_e$ indicates a slight increase (approximately $20-30 \%$) since $z=1$ \citep[e.g.,][]{vanderwel2014, mowla2019,Nedkova2021, Kawinwanichakij2021}. However, this mild size evolution conflicts with theoretical predictions that predict that disk galaxies should have grown approximately by a factor of two in size since z$\sim$1 \citep{mo1998size-simu}, a trend that has also been confirmed observationally using other size indicators \citep{buitrago2024, Yang2025}. 

One promising alternative is to examine galaxy size evolution through a more physically motivated size estimation. \samane{In this work, we use the one proposed by \cite{Trujillo2020} and further developed in \cite{chamba2022}, which is based on the gas density threshold for star formation. This threshold should leave a truncation imprinted on the stellar mass density profile. The radial location of this truncation, R$_{edge}$, coincides with the outermost position of in situ star formation. Therefore, the location of R$_{edge}$ should reflect either the end of present star formation or the location up to which in situ star formation took place in the past.}

Using this novel size proxy, \cite{buitrago2024} analysed 1048 disk galaxies with stellar masses M$_{\star}$ $\sim5\times10^{10}$ \(\textup{M}_\odot\) up to $z=1$ in the HST CANDELS fields. They showed that, at a fixed stellar mass, MW-like disk galaxies have increased in size by a factor of two in the past 8 Gyr. Therefore, MW-like disk galaxies have actually experienced a significant growth in size over time, more in line with theoretical expectations \citep{mo1998size-simu}. Nevertheless, the mechanisms that drive this growth remain unclear. The main question is whether this increase in size \samane{is driven by internal processes, such as in situ star formation or stellar migration, by external mechanisms, such as gas infall or mergers with other galaxies, or by a combination of both}.

The predominant view is that galaxies evolve from the inside out, with the central regions forming first and subsequent star formation or stellar accretion gradually extending to the outer parts over time.  \cite{peebles1969origin} already suggested that gas with lower angular momentum would settle and form stars earlier than gas with higher angular momentum. Both semi-analytic models and cosmological simulations support this theory, demonstrating that star formation tends to occur later at larger radii due to the accretion of gas with higher angular momentum \citep[e.g.,][]{fall1980formation,mo1998size-simu,somerville2008formation,dutton2011evolution,Brook2006,Aumer2014,Avila-Reese2018}. 
Other proposed scenarios that could cause inside-out growth include minor mergers, which add stars to the outskirts of a galaxy \citep[e.g.][]{Naab2009,Hilz2013, Lagos2018}, or stellar migration, where stars formed in the inner regions of a galaxy move outward and contribute to the growth of the outer disk. \citep{Sellwood2002,roskar2008b, Vera-Ciro2014}. 

In observations there is the general agreement that star formation is suppressed in their inner regions compared to their outer regions at $z\sim1$ \citep[e.g.,][]{nelson2016, nelson2019, Morselli2019, Nelson2021, Roper2023}, indicating inside-out growth. However, there is an ongoing debate on the evolutionary path of MW-like galaxies \citep{vanDokkum2013, Patel2013, Tan2024}.

The goal of this study is to analyze MW-like disk galaxies at $z\sim1$. We want to investigate whether the growth of these galaxies is primarily driven by inside-out processes and to determine which mechanisms are responsible. Different growth scenarios leave distinct signatures in stellar populations across galactic radii. For instance, in the case of minor mergers, older stars dominate the central regions formed early in the galaxy’s history, while accreted satellite galaxies — typically old, quenched, and metal-poor due to their lower mass and metallicity — deposit their stars in the outskirts \citep{Naab2009, Oser2010}. This will show as a flatten metallicity profile and a flat, or even positive, age profile \cite{Kobayashi2004,DiMatteo2009, Yoon2023}. Alternatively, stellar migration redistributes stars from the inner disk to larger radii, causing the reshaping of the stellar age radial profile into a distinctive U-shape \citep{Roskar2008}. Meanwhile, the metallicity profile generally exhibits a decreasing trend with radius \citep{Ho2015,Zihao2025}.

Therefore, exploring age and metallicity gradients in galaxies may help us to understand how galaxies have grown and evolved. By linking this information to R$_{edge}$, we can gain a more complete picture of the processes that have shaped their mass assembly over cosmic time.

One of the main challenges in observing distant galaxies is the cosmological dimming, which makes it difficult to observe their outskirts. Additionally, these galaxies appear smaller in the sky, which requires high spatial resolution to be able to resolve them. However, JWST and HST data have significantly improved our ability to observe distant galaxies due to their sub-arcsec spatial resolution from the UV to the NIR wavelengths. Moreover, state-of-the-art methodologies have been developed to accurately remove the effects of the PSF \citep[see e.g.][]{2016ApJ...823..123T,2020MNRAS.491.5317I,2021A&A...654A..40T,2025OJAp....8E..73S,Golini2025} and provide deeper insights into the mechanisms driving galaxy growth.

This paper is structured as follows. Section \ref{sec:samaple} describes the datasets we used and the selection criteria for our galaxy sample. Section \ref{sec:methodology} explains the methodology we applied to derive the radial profiles of the galaxies. The results are presented in Section \ref{sec:result}. In Section \ref{sec:discussion}, we discuss these results, and Section \ref{sec:summ} summarizes our main conclusions. Our adopted cosmology is $\Omega_m =0.3$, $\Omega_\Lambda =0.7$ and $H_0 = 70\, \mathrm{km} /{\mathrm{s/Mpc}}$. All magnitudes in this paper are in the AB system \citep{Oke&Gunn}.

\section{Sample selection and data}\label{sec:samaple}
\subsection{Sample} 
This study aims to present a pilot study on the stellar populations of MW-like galaxies, particularly their outer parts, at $z\sim1$ to gain insight into the processes that drive galaxy evolution. For this reason, our sample is composed of two low inclination, disk galaxies, UDF 5417 and UDF 3372, with a spectroscopic redshift of $z\sim$1 and stellar mass of $M_\star \sim \samane{4\times}10^{10} \, M_\odot$. \samane{The stellar masses used throughout this work are derived from the integration of the stellar mass surface density profiles up to R$_{edge}$ (see Section \ref{sec:mass-profile} for details). We refer to these galaxies as MW-like disk galaxies because they have stellar masses similar to the Milky Way.} These two galaxies are in the Hubble Ultra Deep Field \citep[HUDF, ][]{HUDF}  footprint, one of the best studied fields in Extragalactic Astronomy, and were previously studied in \citet{Trujillo2013}. The coordinates, redshift, and physical scale of the galaxies are given in Table \ref{table1}, which are taken from \citet{Trujillo2013}. 
Here, we will present the methodologies developed that will be applied in larger samples in future studies.

\subsection{JWST Data}
To be able to study the edge of galaxies at $z\sim1$, it is necessary to have both ultra-deep imaging and superbe spatial resolution. Consequently, we used the deepest optical and near-infrared imaging ever carried out with the Hubble Space Telescope (HST) in the GOODS South field, complementing it with ultradeep data from the James Webb Space Telescope (JWST). We use the reduced JWST Advanced Deep Survey (JADES) (PID: 1880, PIs: Daniel Eisenstein, Nora Luetzgendorf) Public Data Release v2.0\footnote{\href{https://archive.stsci.edu/hlsp/jades}{https://archive.stsci.edu/hlsp/jades}} \citep{Rieke2023}. This release covers approximately 25 arcmin$^{2}$, with exposure times per filter ranging from $\sim$14,000 to $\sim$60,000 seconds. The dataset provide imaging in seven wide-band filters (F090W, F115W, F150W, F200W, F277W, F356W, and F444W) and seven medium-band filters (F182M, F210M, F335M, F410M, F430M, F460M, and F480M), all with a pixel scale of 0.03\arcsec\ per pixel. Note that UDF3372 was not observed in 3 medium bands: F430M, F460M, and F480M.

\subsection{HST Data}
Additionally, we used the HST imaging of the Hubble Legacy Field (HLF) v2.0 release \citep{hlf1}, which was retrieved directly from repository\footnote{{\href{https://archive.stsci.edu/prepds/hlf/}{https://archive.stsci.edu/prepds/hlf/}}}. This includes all the 
optical (ACS/WFC F435W, F606W, F775W, F814W, and F850LP) and infrared (WFC3/IR F098M, F105W, F125W, F140W, and F160W) imaging. The total HST on-target exposure time of all the data included in the HLF-GOODS-S is 6.4 Msec. The images used here are drizzled science images with a pixel size of 0.06\arcsec\ . 

\subsection{PSFs}

The primary effect of the Point Spread Function (PSF) of an image is to redistribute light from the inner parts of objects to the outer parts, effectively smoothing the light profile of galaxies and altering their properties. This smoothing effect, which at first order depends on the wavelength, affects the \samane{amount of light beyond} the edge of galaxies. This means that we need to correct for it. To do this, we use a set of PSFs that are used to correct for this smoothing effect in each individual band.

For the HST data, we construct empirical PSFs from stars within the same field of view, \samane{independently for each bands}. For this, we visually select isolated stars to avoid contamination by nearby sources. After removing the sky background, which is explained in Section \ref{sky-bg}, the cut-outs are normalized to 1.

For JWST imaging, we simulated the PSFs using the STScI PSF Simulation Tool\footnote{\href{https://www.stsci.edu/jwst/science-planning/proposal-planning-toolbox/psf-simulation-tool}{STPSF tool}}, which is based on the WebbPSF software \citep{STPSF}\footnote{Previous studies have found that the PSFs derived with WebbPSF are narrower than point sources. However, the discrepancy is only of 1-2\% \citep{Weaver2024}. Therefore, we decided to use the synthetic PSFs because of their higher SNR}. This choice was motivated by the complex shape of the JWST PSF, which has bright diffraction spikes and extended features that make it difficult to find clean, uncontaminated stars for building an empirical PSF. \samane{PSFs are created independently for each band.}

In order to ensure the accuracy of deconvolution, it is crucial that the PSF image extends well beyond the size of the target galaxy. Consequently, for both JWST and HST, we created PSF cut-outs with radii of approximately 2 arcseconds, equivalent to around 1.5 times the mean size of a galaxy in our sample \citep{Sandin2014,Sandin2015}.

\begin{table}[]
\caption{Summary of the main properties of the two target galaxies.}
\renewcommand{\arraystretch}{1.5}  
\label{table1}
\resizebox{\linewidth}{!}{%
\begin{tabular}{|c|c|c|c|c|c|}
\hline
Name    & \begin{tabular}[c]{@{}c@{}}R.A. \\ (J2000)\end{tabular} & \begin{tabular}[c]{@{}c@{}}Decl.\\ (J2000)\end{tabular} & Redshift &  \begin{tabular}[c]{@{}c@{}}Physical scale \\ (kpc/arcsec) \end{tabular}& \begin{tabular}[c]{@{}c@{}}Stellar mass\\ (M$_\odot$)\end{tabular} \\ \hline
UDF 3372 & 03$^{h}$ 32$^{m}$ 42.3$^{s}$  & -27$^\circ$\ 47$^{'}$ 46$^{''}$  & 0.996    & 8.001      & $3.7\times10^{10}$  \\ \hline
UDF 5417 & 03$^{h}$ 32$^{m}$ 39.9$^{s}$  & -27$^\circ$\ 47$^{'}$ 15$^{''}$  & 1.095    & 8.166      & $3.9\times10^{10}$  \\ \hline
\end{tabular}%
}
\tablefoot{The values are adopted from \citet{Trujillo2013}, except for the stellar mass, which has been estimated in this work.}
\end{table}

\section{Methodology}\label{sec:methodology}

To explore the stellar populations of galaxies at $z=1$, we need to process the data in a specific way to minimize biases in our results. In this section, we explain the steps taken to process the observations and to derive the stellar population gradients.

\begin{figure*}[h]
    \centering \includegraphics[width=\linewidth,height=0.66\linewidth]{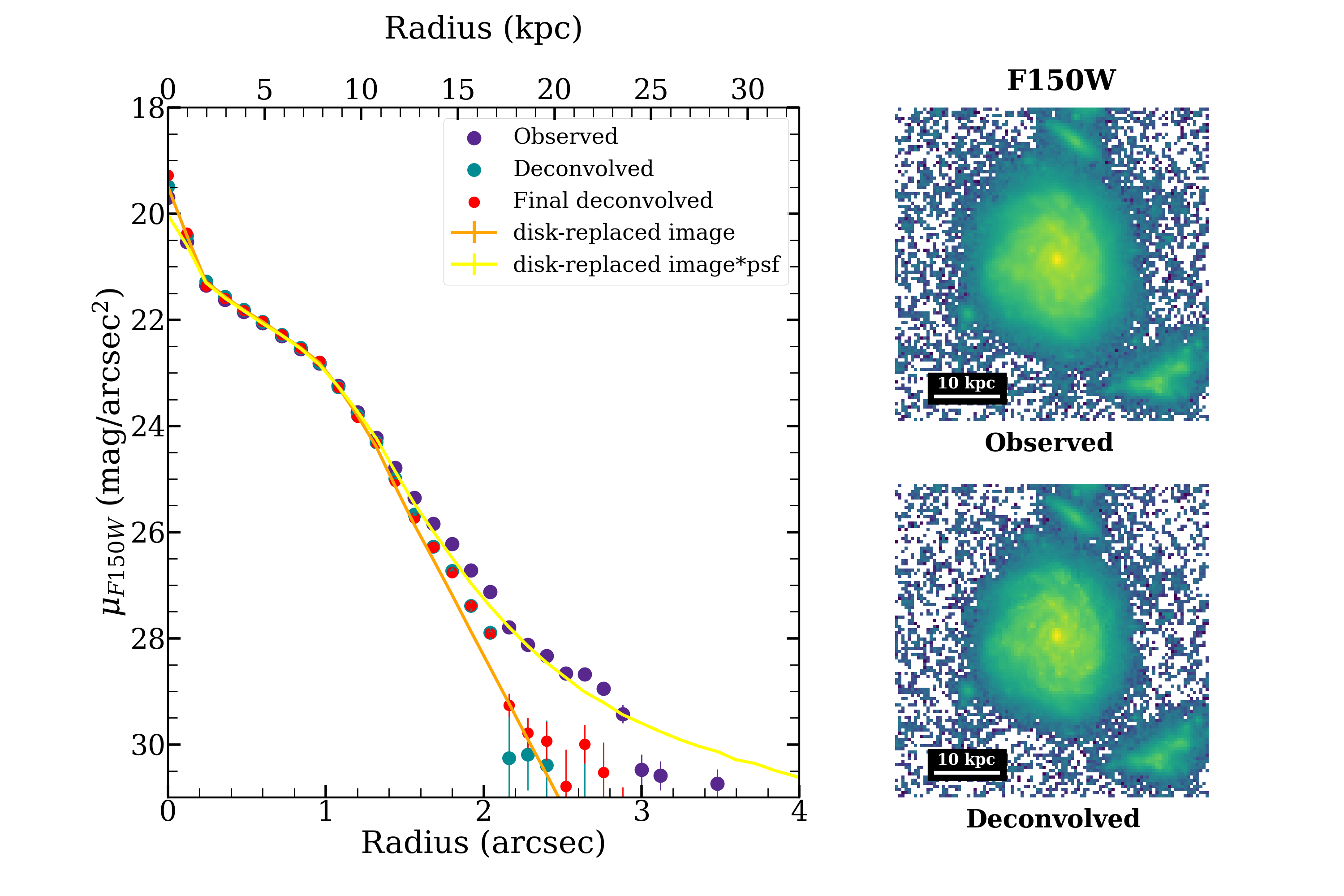}
    \caption{Left panel: Surface brightness profiles of the galaxy UDF 5417 for the F150W band. This figure illustrates the different steps in the deconvolution process. We start with the observed galaxy (purple circles) and deconvolve it using the Wavelet deconvolution (teal circles). \samane{We then replace the inner parts of the model with the galaxy deconvolved (disk-replaced image, orange line) and convolve it with the image's PSF (yellow line)}. The final step is to obtain the residuals (observed - convolved disk-replaced image) to add them back to the model galaxy to obtain the final deconvolved image of the galaxy (red circles).
    Right panels: The galaxy UDF 5417 in the F150W band. The top panel shows the observed image of the galaxy, and the bottom panel shows the final deconvolved image.}
    \label{psf_corr}
\end{figure*}

\subsection{Masking}
Deep imaging of the faint outskirts of galaxies requires careful handling to minimise the light contamination from foreground and background objects that could impact the light profile of the galaxies we want to study.
It is therefore necessary to mask the neighbouring objects carefully. We use the JWST F356W filter as our detection image. Although the F356W filter has lower spatial resolution than filters at shorter wavelengths, it is more sensitive to redder, more evolved or dust-obscured galaxies, particularly at intermediate redshifts. Its longer wavelength also allows the detection of faint extended sources that may be missed in shorter bands.

We use \texttt{NoiseChisel} \citep{gnuastro} to construct segmentation maps. \texttt{NoiseChisel} is a noise-based, nonparametric detection software specifically designed to identify diffuse or low surface brightness structures in deep imaging data. It identifies groups of connected pixels - known as "clumps" - which are then merged to form the final segmentation map. This segmentation image is then converted into a mask, leaving the target galaxy unmasked. Additional manual masking is also necessary for objects that are very close to the main galaxy and have not been identified as separate clumps by \texttt{NoiseChisel}.

\subsection{Measuring the local sky background}\label{sky-bg}

Accurately estimating the local sky background is crucial for determining reliable galaxy surface brightness profiles. To this end, we follow the method described in \cite{Pohlen_2006}. Briefly, after masking all detections except the main galaxy, we compute its intensity radial profile by measuring the average flux in elliptical annuli centered on the galaxy. Examining the radial profile allows us to identify the region where background noise dominates over the galaxy signal, as indicated by a flattening of the profile. We then estimate the sky background by taking the median flux value in this outer region and subtracting it from the entire image.

\subsection{Surface brightness radial profiles}

We compute the surface brightness radial profiles using elliptical annuli as well, but converting the flux values to $mag/arcsec^{2}$. We use the \texttt{astscript-radial-profile} tool from GnuAstro \citep{Infante-Sainz2024} to generate 1D elliptical radial profiles from the 2D galaxy image. For each galaxy, we estimated visually the axis ratio and position angle.

The surface brightness radial profiles for all of the filters are shown in Fig.~\ref{sb-prof} in Appendix \ref{app:sb_prof}. 
We estimate the error in each radial bin in the following way. First, we measure the sky noise by calculating the Root Mean Square (RMS) in an empty region of the image. Then, we scale it by the number of pixels in each bin. We compute the total uncertainty as the combination of the Poisson noise in each bin and the sky error.

The location of the edge, as found by \cite{chamba2022}, can be seen as a sharp drop in the surface brightness profiles, particularly noticeable in the blue filters. However, the observed surface brightness profiles are affected by several factors that can alter the characterisation of this edge, including inclination, Galactic extinction, cosmological dimming, and the PSF. While the first three affect the entire radial profile, the effect of the PSF is to change the intrinsic shape of the galaxy, meaning that it varies at different radii.

\subsection{PSF correction}\label{sec:psf}

The advantage of space-based imaging is the improved resolution and sharpness of the PSF when compared with ground-based imaging. However, the effect of the PSF of the telescope is still significant for smaller, more compact objects, such as those examined in this study. Therefore, we need to correct for the PSF effect and deconvolve our galaxy. However, deconvolution algorithms do not preserve the noise characteristics of the images. Consequently, we developed a method similar to that of \cite{Golini2025}, to deconvolve the images from the PSF while preserving the signal-to-noise ratio. The steps performed are as follows and are illustrated in Fig.~\ref{psf_corr}.

\begin{enumerate}
    \item The first step is to rotate the JWST PSFs by 60 degrees in order to match them with the stellar spikes in the images. In the case of the HST, the PSFs are left untouched, as we are using PSFs directly taken from the data.
    
    \item To perform the deconvolution, we used the Wavelet deconvolution algorithm. Wavelet deconvolution methods employ spatial regularization based on the assumption of sparsity in the wavelet domain \cite[][]{Starck1994, Starck2006}. This approach effectively preserves image features across multiple spatial scales; however, it may suppress low surface brightness structures. Consequently, careful treatment of galaxy outskirts is required to prevent the introducing of artifacts.
    We apply the wavelet deconvolution method\footnote{The code is provided in \href{https://github.com/aasensio/wavelet_deconvolution?tab=readme-ov-file}{https://github.com/aasensio/wavelet-deconvolution}} using the corresponding PSF for each image to obtain a deconvolved image and the corresponding surface brightness radial profile (shown in teal circles in Fig.~\ref{psf_corr}). 

    \item  We then use IMFIT \citep{Erwin2015} to create a model of the galaxy based on the deconvolved image, modelling both the bulge and disk components. The bulge is modelled using a 2D S\'ersic, typically with a fixed index of  $n=1$. However, we allow $n<1$ in some filters, as this provides a better representation of the observed light distribution in those cases.
 For the disk component, IMFIT allows us to use the \texttt{BrokenExponentialDisk3D} to model the disk of the galaxy and its edge.   
    
    \item \samane{After determining the best-fitting model for each galaxy, we replace the central region of the model, within 1\arcsec\ for UDF 3372 and 1.3\arcsec\ for UDF 5417, with the corresponding pixels from the deconvolved image.} These radii have been chosen so as to encompass the discs of the galaxies. We call this image a disk-replaced image (orange line). 
    
    \item Using this disk-replaced image and convolving it with the PSF, we are able to reproduce the observed galaxy (yellow line).
 
    \item The convolved disk-replaced image is subtracted from the observed image to obtain the residuals of the image. These residuals contain asymmetries in the galaxy that are difficult to model with IMFIT and, more importantly, the sky background of the image, but with the effect of the PSF removed. 
    
    \item Finally, the residuals are added to the disk-replaced image, resulting in a deconvolved image of the galaxy that effectively preserves the signal-to-noise of the sky background (red circles). 
\end{enumerate}

To test the robustness of the wavelet algorithm, we also used the Wiener deconvolution algorithm in \texttt{scikit-learn} \citep{scikit} instead, which gives comparable results. The results of this test are shown in the Appendix \ref{app:wiener}.

\subsection{Inclination correction}
The inclination of the object strongly affects the surface brightness profiles of disk galaxies; the more inclined the object is, the brighter it appears (see figure 2 in \cite{Martin-Navarro2014}). To correct for this effect, we use the equation provided in \cite{Trujillo2020}:

\begin{equation}
\mu_{\mathrm{corr}} = \mu_{\mathrm{observed}} + \sum_{j=0}^{4} \alpha_j \left( \frac{b}{a} \right)^j
\end{equation}

where $b/a=cos(i)$ is the axis ratio of the galaxy, being $i$ the galaxy's inclination. In \citet{Trujillo2020}, the $\alpha_j$ coefficients are calculated assuming different disk thicknesses, characterized by the ratio $z_0/h$, where $h$ is the disk scale length and $z_0$ the scale height. In this work, we assume $z_0/h = 0.12$, corresponding to a moderately thick disk, \samane{for all bands. As the galaxies in our sample are nearly face-on, uncertainties in the $z_0/h$ parameter are expected to have a minimal impact on our results. Variations in $z_0/h$ change the surface brightness by at most 0.005–0.008 mag, corresponding to 0.025–0.04$\%$.}

\subsection{Galactic extinction and cosmological dimming correction}

To obtain the final radial surface brightness profiles of our galaxies, we need to correct for Galactic extinction and cosmological dimming.
For the Galactic extinction, we used the values provided by the NASA/IPAC Extragalactic Database (NED) Extinction Calculator\footnote{\href{https://ned.ipac.caltech.edu/extinction_calculator}{\texttt{https://ned.ipac.caltech.edu/extinction\_calculator}}} \citep{Schlafly2011}.
 The values used for each galaxy and filter are given in Table \ref{table_ext}. The NED extinction calculator only provides extinction values for the HST filters. For JWST, the extinction values for F090W, F115W, and F150W were taken from the closest HST filter. For the reddest filters (F200W, F277W, F356W, F444W, F182M, F210M,
F335M, F410M, F430M, F460M, and F480M), we set the extinction to 0, as these are the least affected and the extinction is already low (0.004) at F160W.

\begin{table}[!h]
\caption{Galactic extinction for the two galaxies in our sample in each of the observed HST and JWST filters.}
\renewcommand{\arraystretch}{1.5}  
\resizebox{\linewidth}{!}{%
\label{table_ext}
\begin{tabular}{|c|c|c|c|c|c|c|c|c|}
\hline
Galaxy  & F098M & F090W & F105W & F115W & F125W & F140W & F150W & F160W \\ \hline
UDF 3372 & 0.007 & 0.007 & 0.007 & 0.006 & 0.006 & 0.005 & 0.004 & 0.004  \\ \hline
UDF 5417 & 0.008 & 0.008 & 0.008 & 0.006 & 0.006 & 0.005 & 0.004 & 0.004 \\ \hline

\end{tabular}%
}
\end{table}

To correct for cosmological dimming, we used the following expression:
\begin{equation}
    \mu_{\text{corr}} = \mu_{\text{observed}} - 7.5 \times \text{log}_{10}(1+z)
\end{equation}

because we are working with flux densities, as explained in \citet{buitrago2024}.

Fig.~\ref{decon-prof} in Appendix \ref{app:dec_prof} shows the final surface brightness profiles of the two galaxies with all corrections in all filters. Note the difference between the observed (Fig.~\ref{sb-prof}) and deconvolved profiles, especially in the outer regions. This shows that a significant fraction of the light in these outer regions is produced by the PSF effect, which highlights the importance of correcting this effect to obtain credible results.

\subsection{Spectral Energy Distributions}

\begin{figure*}[h]
    \centering \includegraphics[width=1\linewidth,height=1\linewidth]{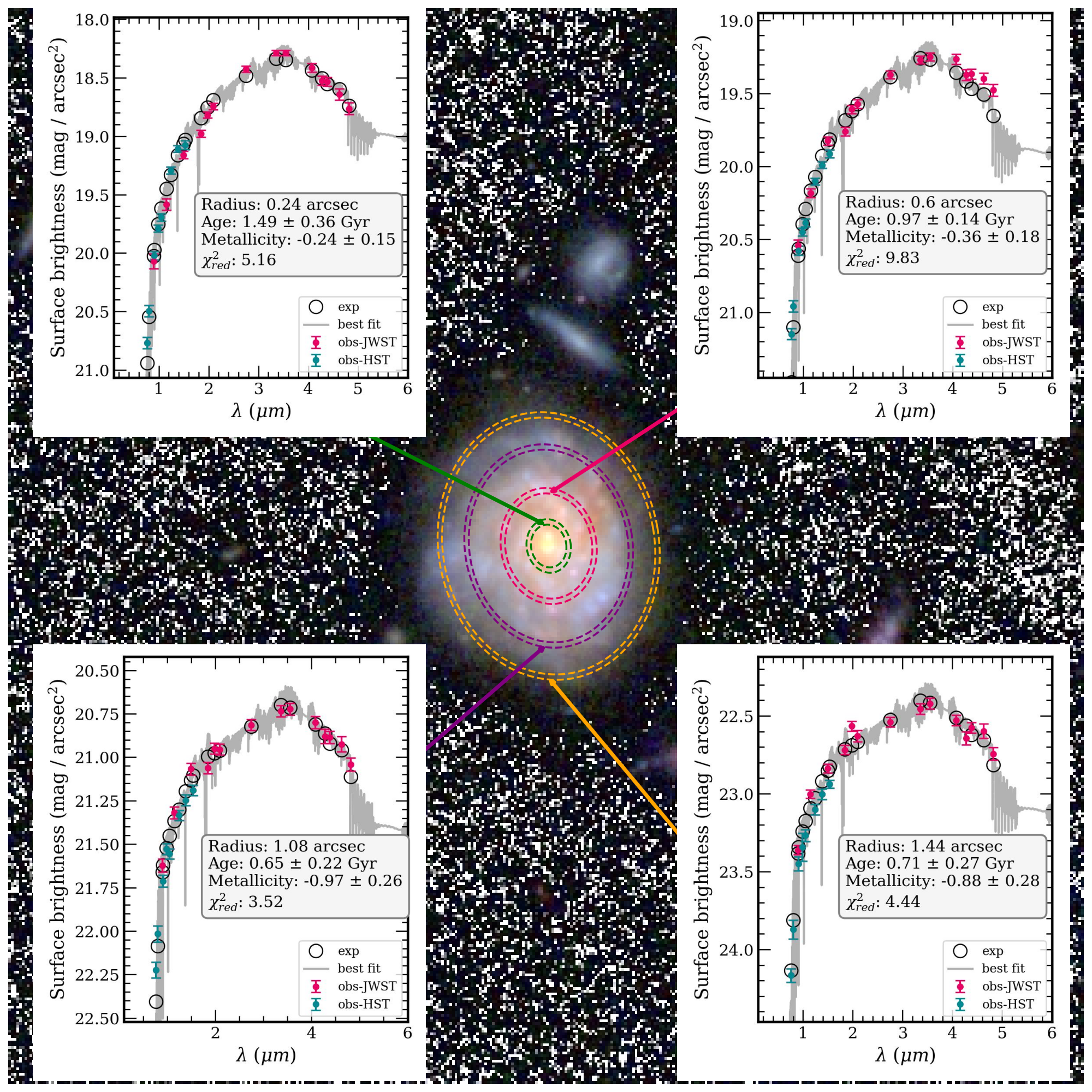}
    \caption{Example of 4 SEDs at different radial distances for the galaxy UDF 5417. The figure shows the SEDs from a region close to the central core (top left) to the outer regions (bottom right), as indicated by the colours of the ellipses. In each panel, the blue and red circles represent the photometry of the galaxies (HST and JWST filters, respectively). The grey continuum is the best-fit SSP model, and the open circles indicate the convolution of the model with the response of the filters used here. The reduced $\chi^2$, estimated age, metallicity and distance are reported in each plot. }
    \label{SEDfig}
\end{figure*}

The main goal of this work is to study the stellar population properties of two galaxies, from their central regions to well beyond their outskirts, to understand their evolution with time. Taking advantage of the broad wavelength coverage of our photometric dataset, we fit single stellar population (SSP) models to derive radial gradients in stellar age and metallicity.

Our analysis follows the methodology described in \citet{Montes2014} and \citet{MT18}, as summarized below. To characterize the stellar populations at different galactocentric distances, we compare the observed spectral energy distributions (SEDs) with a grid of SSP models. First, the models are redshifted to match the galaxy's redshift. These redshifted spectra are then convolved with the filter response of the observed photometric filters to generate synthetic photometry. We perform a $\chi^2$ minimization (see eq. 2 in \citealt{Montes2014}) to identify the best-fitting model and estimate the 1$\sigma$ confidence intervals. The three free parameters in the fitting process are stellar age, metallicity, and luminosity. Therefore, this SED-fitting procedure enables us to estimate the \samane{light-weighted }age and metallicity profiles as a function of radius for each galaxy.

We use the UV-extended E-MILES SSP models \citep{vazdekis2016}, which are based on Padova 2000 isochrones \citep{Girardi2000}. These models span a wavelength range of 1680-50\,000 \AA{}. The original model grid includes seven metallicities (-1.79 $\leq$ [Fe/H] $\leq$ 0.26) and 50 ages ranging from 0.03 to 17.8 Gyr. To improve the accuracy in metallicity, we expanded the grid of models with 200 metallicities, linearly interpolating the original SSPs. \samane{Interpolating the original model grid improves the accuracy of the metallicity estimate when the SED changes smoothly. However, there are ranges of ages and metallicities where the SED can experience significant local flux variation at specific wavelengths. In these cases, linear interpolation does not necessarily improve the accuracy with which the metallicity can be determined.}

\samane{ Since our galaxy sample is located at redshift $\sim$1, we restrict the maximum allowed SSP age to that of the Universe at the galaxy's redshift. This ensures that the stellar population models remain physically consistent with cosmological age constraints.}
Our choice of IMF is a Kroupa Universal IMF \citep{Kroupa2001}. 

Note that considering an SSP at each radius is a rough assumption, since we expect the star formation history of our galaxies to be more complex. However, this approximation is acceptable as we are interested in the average age and metallicity. Additionally, the two bluest optical filters (F435W and F606W) were excluded from the SED fitting due to their extremely low signal-to-noise ratio in the outer regions of the galaxies.

Fig.  \ref{SEDfig} shows 4 examples of SEDs at different radii for one of our galaxies, UDF 5417. It also shows the best-fitting model (grey solid line) with the open circles being the convolution of this model with the response of the HST and JWST filters.

\begin{figure}[]
    \centering
    \subfigure[UDF 3372]{
        \includegraphics[width=0.5\textwidth]{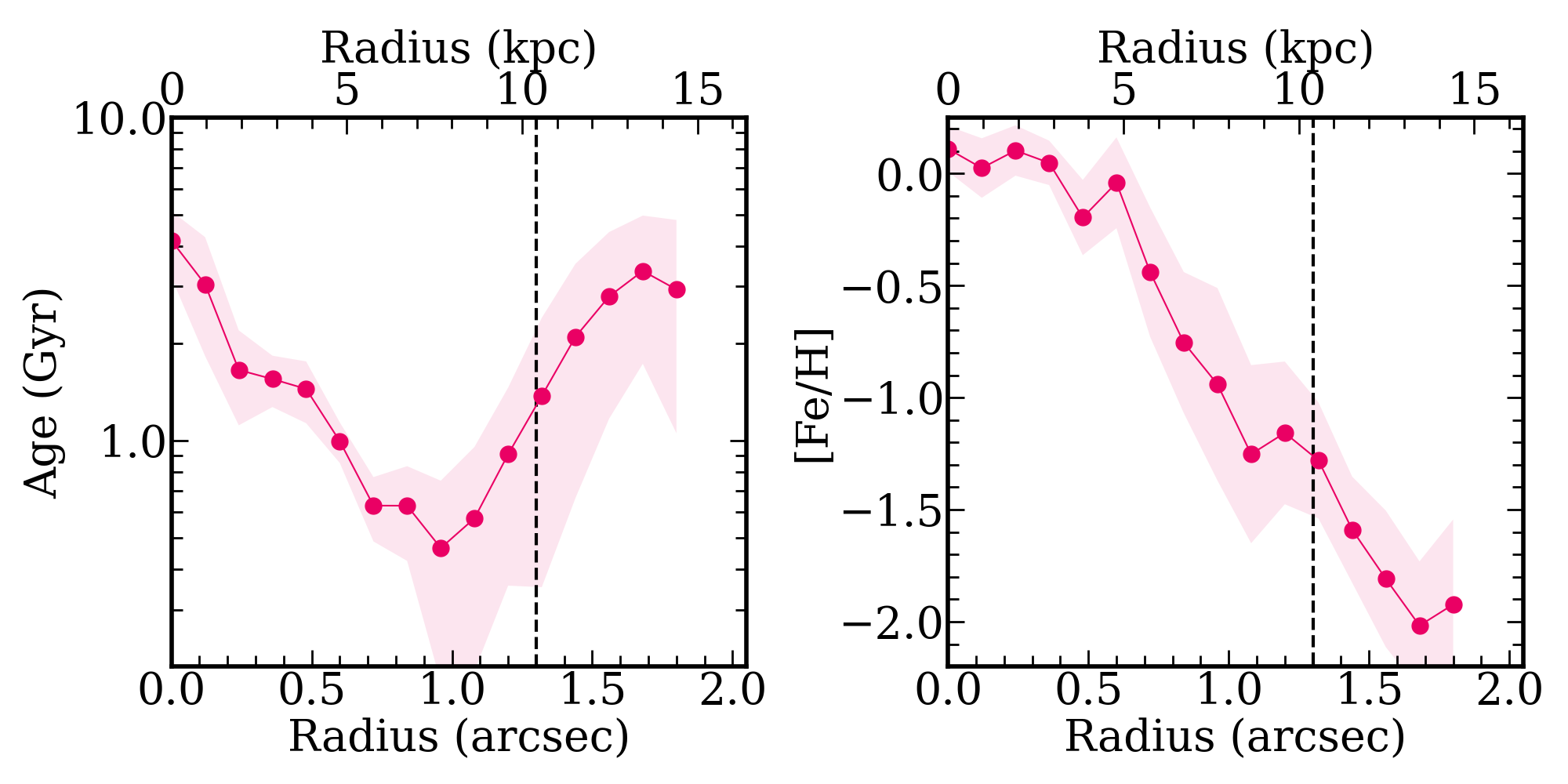}
    }
    \subfigure[UDF 5417]{
        \includegraphics[width=0.5\textwidth]{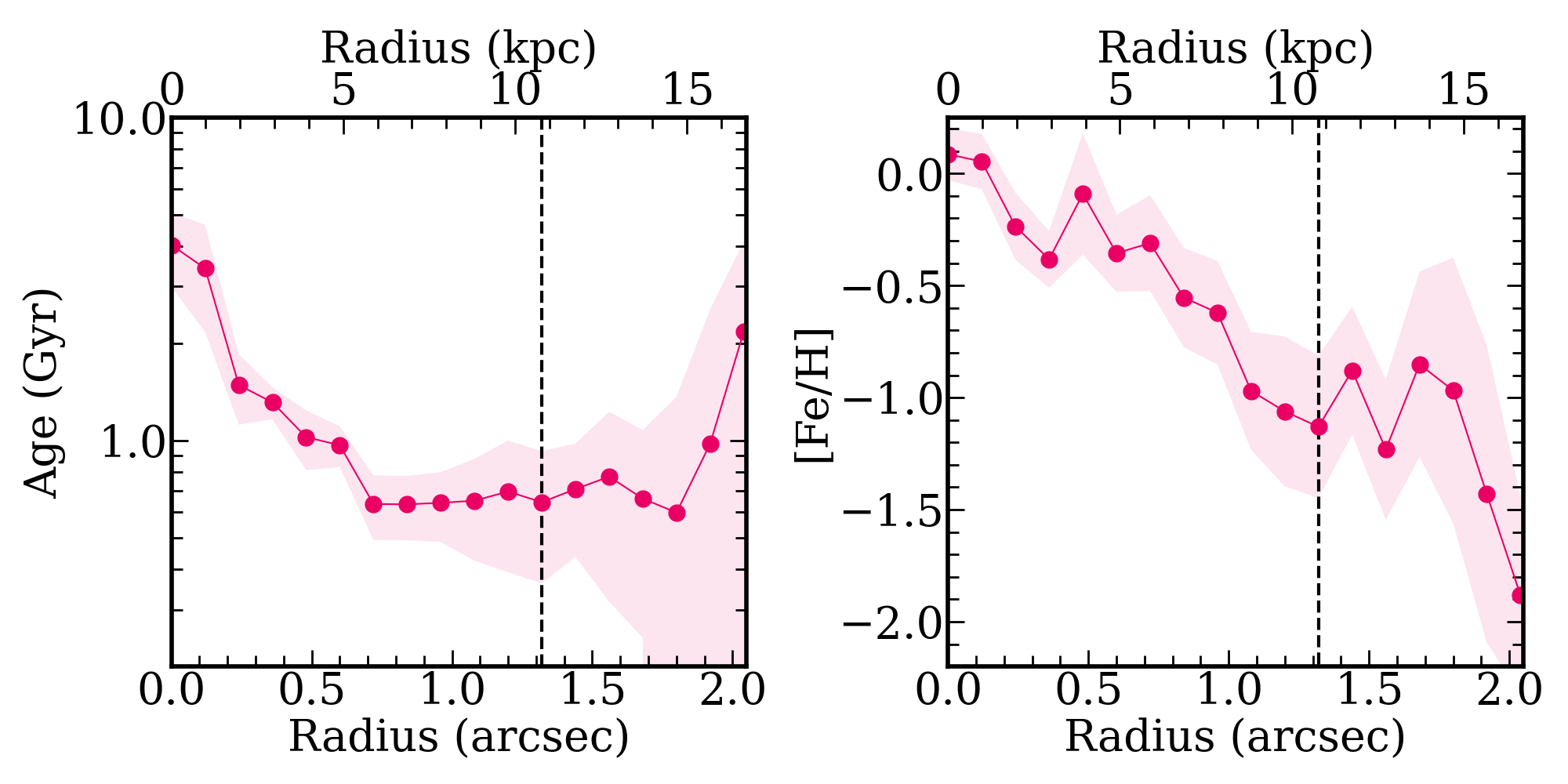}
    }    
    \caption{Age and metallicity profiles for the galaxies UDF 3372 (top panels) and UDF 5417 (bottom panels). The left panels show the age radial profiles, while the right panels show the metallicity profile. The 1$\sigma$ uncertainties are indicated by the pink shaded regions. The dashed line denotes the location of the galaxy edge, R$_{edge}$.}
    \label{age-met}
\end{figure}

\begin{figure}[h]
    \centering
    \subfigure[UDF 3372]{
        \includegraphics[width=0.5\textwidth]{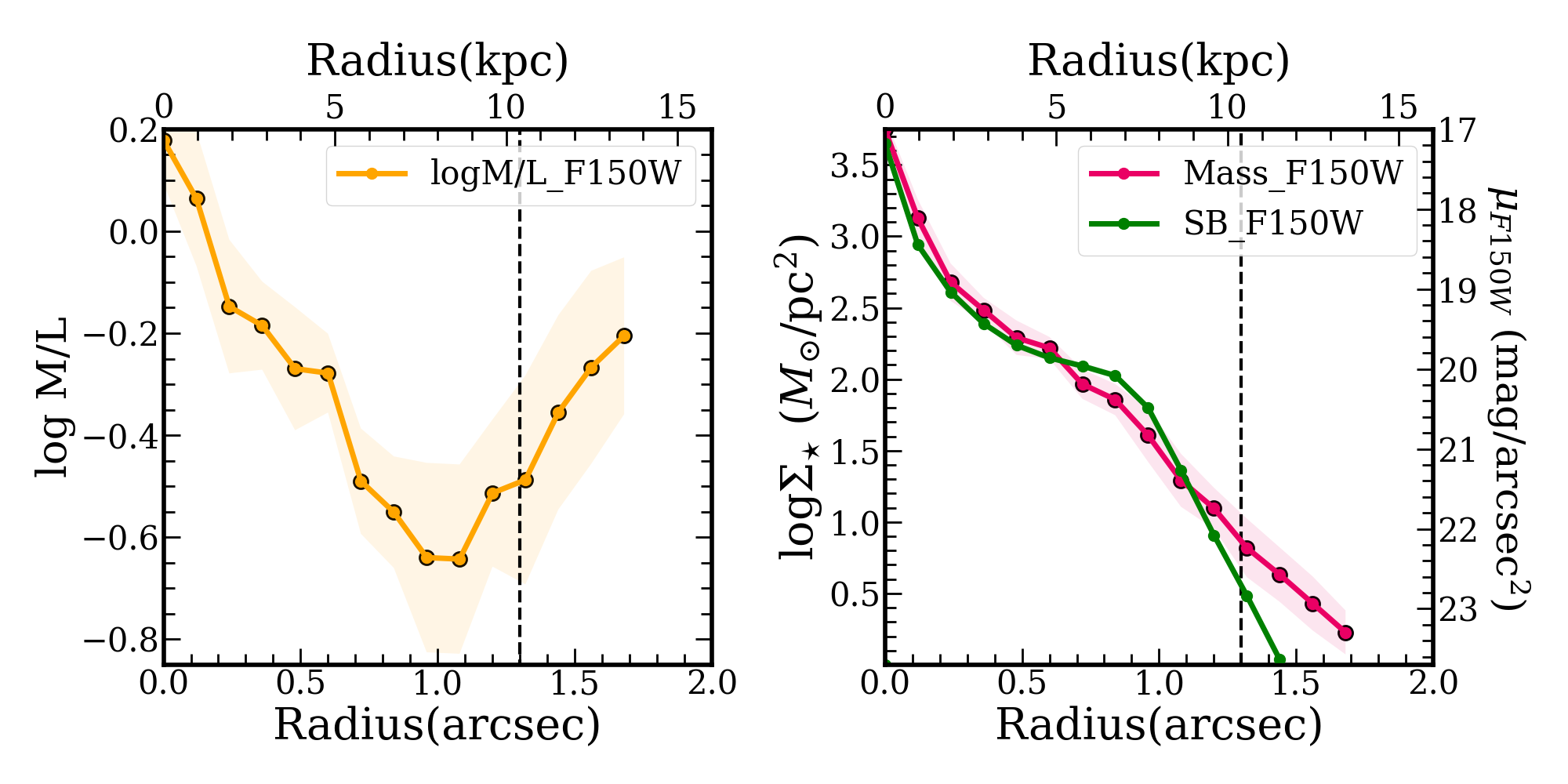}
    }    
    \subfigure[UDF 5417]{
        \includegraphics[width=0.5\textwidth]{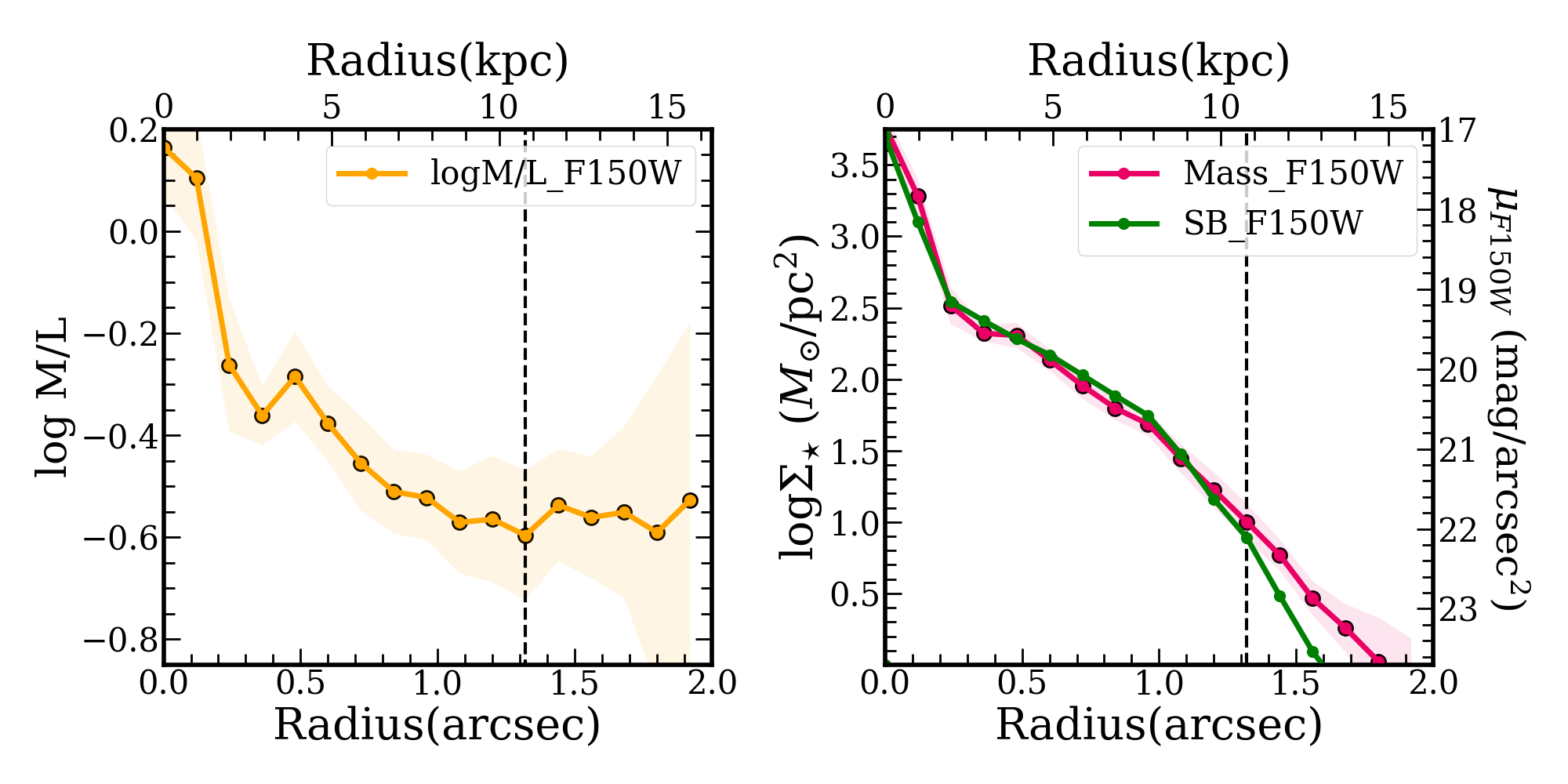}
    }
    \caption{ M/L and stellar mass density profiles for the galaxies UDF 3372 (top panels) and UDF 5417 (bottom panels). The left panels show the M/L ratio estimated from the stellar populations, while the right panels show the stellar mass density profile (red) and the surface brightness profile in the F150W band.}
     \label{stellar-mass}
\end{figure}

\section{Results} \label{sec:result}

\subsection{The edges of galaxies at z=1}
To identify R$_{edge}$, we examine the surface brightness profile in one of the bluest bands, F775W, after deconvolution, along the wedge direction, searching for a distinct change in slope in the outer regions (see \citealt{chamba2022}). Since galaxies are often asymmetric, wedge profiles are the most reliable way to locate the edge, rather than elliptical apertures \citep{Raji2025}. In parallel, we visually confirm the position of this transition by inspecting the two-dimensional deconvolved FITS images. The R$_{edge}$ for UDF 5417 is 10.6 kpc and for UDF 3372 is 10.4 kpc. Fig.~\ref{decon-prof} shows the surface brightness radial profile of all bands for both galaxies, using elliptical apertures and the location of the edge derived from wedge profiles (vertical red dashed line).

\subsection{Age and metallicity radial profiles}
\label{age-met-sec}

In Fig.~\ref{age-met}, we present the age and metallicity profiles for UDF 3372 (top panels) and UDF 5417 (bottom panels). The age profiles of both galaxies show a distinctive negative gradient, indicating that the stellar populations are younger in regions farther from the galaxy center. Beyond a certain radius, close to R$_{edge}$, the stellar populations become older again. This trend is more evident for UDF 3372, while in the case of UDF 5417, the errors are compatible with a flat age gradient. The metallicity gradients show a decreasing trend with radius, meaning that the stellar populations become increasingly metal-poor with increasing distance to the center.

To verify the reliability of the results obtained here, we performed three different tests to the SED fit: 1) using only the JWST photometry, 2) using the HST and JWST photometry only from wide filters (i.e., removing the medium filters), and 3) applying the PSF correction method described in Sec. \ref{sec:psf}, but using a Wiener deconvolution algorithm instead. More details on the latter are given in the Appendix \ref{app:wiener}, and the corresponding age and metallicity profiles are shown in Fig.~\ref{age-met-wiener}. The results of these three tests are consistent, i.e., they deliver profiles with similar shapes and values to those presented in  Fig.~\ref{age-met}. This indicates that our results are robust to the different filters used and the methodology applied. In addition, we perform the SED fitting using the original observed profiles (before deconvolution) as shown in Fig. \ref{age-met-PSFin}. In this case, we do not recover the trends in age and metallicity, especially the age turnover close to R$_{edge}$. This underlines the necessity to correct for the PSF effect if the stellar populations and other properties of high-redshift galaxies are to be correctly derived.

\samane{The surface brightness profiles used for SED fitting were not corrected for internal dust attenuation. In Appendix \ref{dust}, we evaluate the potential impact of dust attenuation on our results. To estimate this impact, we adopted the model of \cite{Tuffs2004}, which provides wavelength- and inclination-dependent attenuation corrections for spiral galaxies. Since the galaxies in our sample are predominantly face-on, the overall effect of dust is expected to be moderate.
The effects of dust correction on the derived age and metallicity are shown in Appendix \ref{dust-correction-33372}. We also compared the stellar mass profiles with and without dust correction, as shown in Figure \ref{mass-dust-corr}.}

\samane{To evaluate the potential impact of mixing stellar populations when deriving stellar population parameters with SSP models, we conducted an additional test alongside the elliptical analysis. Instead of averaging the flux in ellipses, we measured the surface brightness along several wedges. For each wedge, we calculated separate radial profiles of age and metallicity. We then compared these wedge results to those from the ellipse method. Both methods produced similar radial trends for age and metallicity. Therefore, while we acknowledge that fitting SEDs with SSP models provides a simplified description of stellar populations, this test demonstrates that the overall radial trends reported in this work are robust. This comparison is shown in Appendix \ref{wedge_age_met}.}

We warn the reader that the derived ages and metallicities should be interpreted as SSP-equivalent or luminosity-weighted quantities. Using single SSPs in fitting complex star formation systems can bias results towards younger populations in some cases, or be more affected by age-metallicity degeneracy. Therefore, our methodology is likely more reliable for tracing relative radial trends than for obtaining absolute stellar population parameters.

\subsection{Stellar mass density profiles} \label{sec:mass-profile}

To understand the time evolution of MW-like galaxies, we need to derive the stellar surface mass density profiles for our sample of galaxies. The models of \citet{vazdekis2016} also provide the mass-to-light (M/L) ratios for different photometric bands. After deriving the age and metallicity profiles from SED fitting, we link these ages and metallicities to their M/L at each radial distance. Finally, we calculate the stellar mass density using the following equation taken from \citet{bakos2008color}:

\begin{equation}
\log{\Sigma_*} = \log \left(\frac{M}{L}\right)_\lambda - 0.4 \left(\mu_\lambda - \mu_{\text{abs},\odot,\lambda}\right) + 8.629
\end{equation}

where $\mu _{abs, \odot , \lambda}$ is the absolute magnitude of the Sun in the filter $\lambda$ and $\mu _{ \lambda}$ is the surface brightness in the same filter. To derive the stellar mass density profiles, we use the JWST F150W filter, the reddest filter in the optical rest frame of the galaxies. This filter corresponds to the restframe $\lambda\sim0.75\, \mu m$  for both galaxies and is red enough that it traces the stellar mass of the systems more closely. We estimate the M/L for the rest-frame value of F150W for each galaxy.

Fig. \ref{stellar-mass} shows the M/L (left) and stellar mass density (right) profiles for UDF 3372 (upper panels) and UDF 5417 (lower panels). We also show the F150W surface brightness profiles (right, green lines) of both galaxies for comparison. As expected, the stellar mass density profile follows the F150W surface brightness profile quite closely. However, we see differences between the two profiles, especially in the outer parts, due to the change of trend in the stellar populations outwards R$_{edge}$, as seen in the M/L profiles.

The values of the stellar mass surface density at the edge of the galaxy, $\Sigma_*(\text{R}_{edge})$ are $7.6\pm1.3$ and $11.2\pm1.5$ $M_{\odot}/\text{pc}^2$ for UDF 5417 and UDF 3372, respectively. \samane{The total stellar mass is calculated by integrating the stellar mass density profile, yielding $3.9 \times 10^{10}\ M_\odot$ for UDF 3372 and $4.1 \times 10^{10}\ M_\odot$ for UDF 5417.}

\section{Discussion} \label{sec:discussion}

In this pilot study, we determine the edge radius (R$_{edge}$) of two galaxies at $z=1$. Thanks to the impressive multiwavelength (22 bands) and deep imaging with exquisite spatial resolution provided by HST and JWST, we were able to derive the age and metallicity profiles of these galaxies to distances well beyond their edges ($\sim15$ kpc). Below, we discuss the implications of our findings.

\subsection{Comparison with previous works}

In a recent study, \citet{buitrago2024} explored the evolution of R$_{edge}$ with redshift. They show that galaxies with stellar masses comparable to that of the MW ($\sim5\times10^{10} M_{\odot}$) have increased their size by a factor of approximately two since $z=1$. They estimated that R$_{edge}$ typically lies in the range 10 -- 12 kpc at $z=1$ for those galaxies. For our galaxies, we measured $\text{R}_{edge}=10.6$ kpc for UDF 5417 and $\text{R}_{edge}=10.4$ kpc for UDF 3372, comparable to the sizes reported in \citet{buitrago2024}. Additionally, the values of the stellar mass surface density for our galaxies at their edges, $7.6\pm1.3$ and $11.2\pm1.5$ $M_{\odot}/pc^2$, are in agreement with those in \citet{buitrago2024} ($12.95\pm1.08$ $M_{\odot}/\mathrm{pc}^{2}$). These stellar mass density values are, however, significantly larger than those reported in \citet{martinz-lombilla2019}, \citet{chamba2022} and \citet{Golini2025} at $z=0$ ($\Sigma_*(\text{R}_{edge})\sim1 \ M_{\odot}/\text{pc}^2$). As discussed in \citet{buitrago2024}, this marked evolution of $\Sigma_*(\text{R}_{edge})$ with redshift is likely due to the decrease of the star formation rate over time, allowing progressively lower gas densities to form stars.

\subsection{Interpreting the age and metallicity profiles}
In Section \ref{age-met-sec}, we derived the age and metallicity radial profiles of UDF 3372 and UDF 5417. The metallicity profiles exhibit a negative radial gradient. This is in line with observations and simulations, which consistently show that isolated disk galaxies tend to exhibit negative metallicity gradients \citep[e.g.][]{sanchez2014,Boeche2014, Anders2014, Donor_2020, Renaud2025}. These gradients have been linked to the inside-out growth of galactic disks \citep[e.g.,][]{Matteucci1989, Pilkington2012}.

The age profiles also exhibit a negative gradient, i.e., the ages decrease with radius, until R$_{edge}$, at which point the gradient either remains flat or becomes positive with radius (i.e., older ages with increasing radius). These U-shaped age profiles have also been reported in several previous observational studies \citep[e.g.][]{Azzollini2008,bakos2008color,sanchez2009,Zheng2015,Ruiz-Lara2016,Watkins_2016}. These age gradients are consistent with the results of the N-body simulations in \citet{Roskar2008} that find that a decline in the gas surface density produces a truncation in the star formation (i.e., the gas is not forming stars), which translates into breaks in the stellar mass surface density profiles of galaxies (i.e., an edge). They also showed that stars beyond these edges originate in the disk of the galaxy and migrate radially to populate the outer regions of galaxies \citep[see also][]{Debattista2017}. This radial migration can produce the observed age trend in our galaxies. As we move outwards and the stellar density drops, young stars in the disk progressively dominate the SED. Beyond R$_{edge}$, however, there are almost no young stars, so the SED is dominated by the older, migrated, stars. 

Our data, while easy to accommodate in the stellar migration scheme, it is difficult to fit with other alternatives. For example, tidal stripping of a satellite companion can deposit stars into the outer disc of a host galaxy, introducing older, ex-situ stellar populations to the outskirts \citep[e.g.,][]{Lackner2012,Yoon2023, GonzalezJara2025}. This process can modify the age gradient, either flattening it or producing an upturn in the outer regions, since the accreted stars are typically older than the in-situ populations at similar radii. However, such stellar accretion will also flatten the outer metallicity profiles through the redistribution and mixing of stars \citep[e.g.][]{Kobayashi2004, DiMatteo2009, Cook_2016}. This prediction is in tension with the radial decline of the metallicity we find at all radii.

\begin{figure*}[h]
    \centering \includegraphics[width=1\linewidth,height=0.4\linewidth]{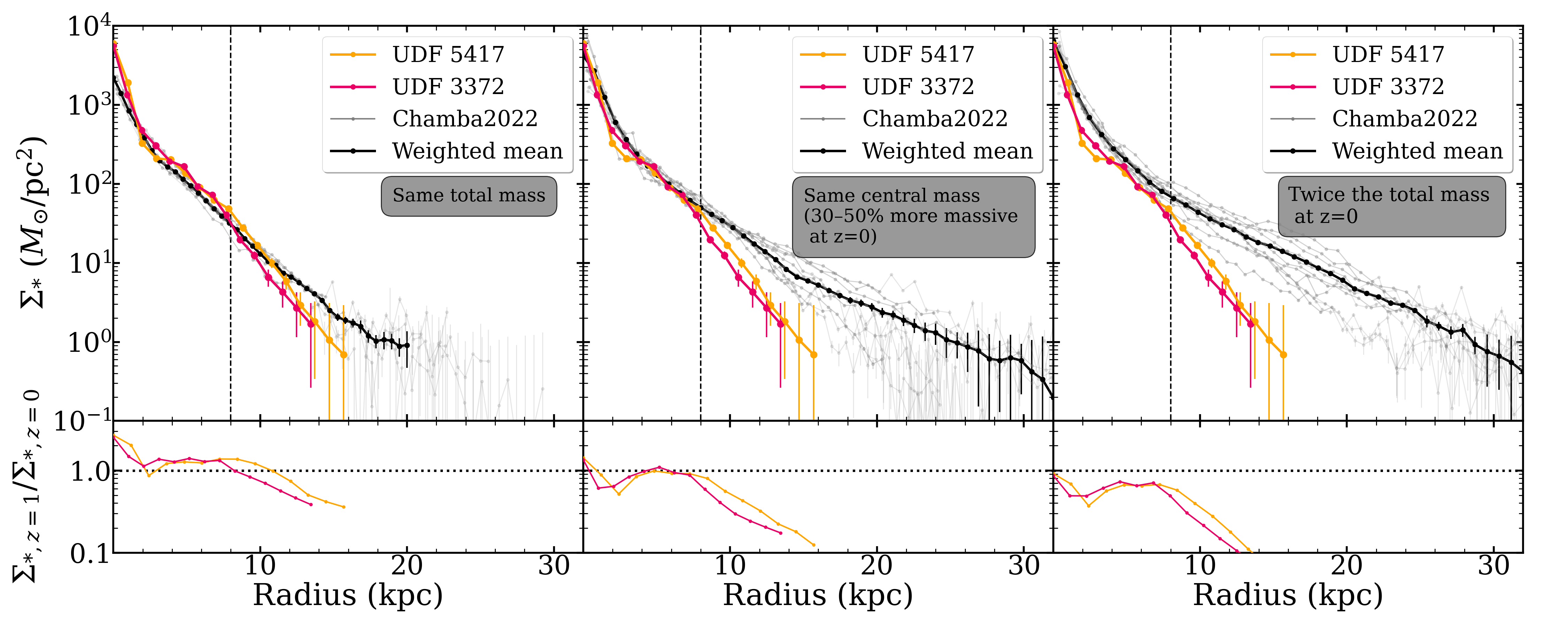}
    \caption{The top panels show the stellar mass surface density profiles ($\Sigma_*$) of the two disk galaxies in our sample at $z=1$ (shown as the yellow and red solid lines) compared to other local disk galaxies from \citet{chamba2022} (grey solid lines). The weighted mean profile of \citet{chamba2022} samples is shown in the black profile. The three panels correspond to different comparison samples at z=0: galaxies with a similar stellar mass (left), galaxies that are approximately $40\%$ more massive (middle), and galaxies that are twice as massive (right). The vertical dashed black line marks 8 kpc, beyond which the mass profiles of the $z=1$ galaxies deviate from those of the $z=0$ galaxies. The lower panels show the ratio of the $z=1$ mass profiles to the weighted mean profile of $z=0$ galaxies, to illustrate the differences at each radius, and the horizontal dotted line indicates where the ratio is equal to 1.}
    \label{stellar-mass-z0}
\end{figure*}

\subsection{The evolution of MW-like galaxies} \label{sec: mass evolution}

Understanding how MW-like galaxies assembled their stellar mass from redshift $z = 1$ to 0 provides key insights into the evolution of galaxies in the latter half of cosmic time. On the one hand, some studies suggest that after $z\sim1$, the growth of the stellar mass of MW-like galaxies slowed substantially, increasing by only about $40\%$ \citep{vanDokkum2013}. On the other hand, other observational studies show that the stellar mass of MW–like star-forming galaxies has increased by a factor of about 2–3 since $z=1$ \citep{Patel2013,Tan2024a,Tan2024,Zasowski2025}.

Based on these considerations, we evaluate three possible evolutionary pathways, illustrated in Fig. \ref{stellar-mass-z0}. These pathways illustrate the potential growth histories of MW–like galaxies over cosmic time: no evolution, moderate growth (30-50\% more massive), and extreme growth, in which the stellar mass nearly doubles. In all cases, we compare the stellar mass density profiles of galaxies in our sample at $z=1$ with those of disk galaxies at $z=0$ from \cite{chamba2022}. To facilitate this comparison, we use the profiles from \citet{chamba2022}. The methodology for calculating the average $z=0$ profiles at each radial distance is detailed in Appendix~\ref{average profiles}.

The left panel of Fig. \ref{stellar-mass-z0} shows galaxies at $z=0$ with the same stellar mass as those from our sample. This plot represents a passive evolution scenario, in which there is no addition of stellar mass and only secular evolution takes place. The ratio of the stellar mass density profiles of our galaxies to the average profiles at $z=0$ (bottom panel) is greater than one from 0 to 8-10 kpc. Since stars in the central region are unlikely to escape the gravitational potential and migrate well beyond the edge of the galaxy, this evolutionary scenario is unlikely, and we do not discuss it further.

In the second scenario, as shown in the middle panel of Fig. \ref{stellar-mass-z0}, we compare our sample with a set of $z=0$ galaxies that are about $30$–$50\%$ more massive than UDF 5417 and UDF 3372, as expected from some previous works \citep[e.g.,][]{vanDokkum2013}. The ratio of the stellar mass density profiles (middle bottom panel) is close to 1 at all radii, but beyond $\sim8$ kpc it starts to deviate significantly, meaning that the $z=0$ galaxies show an excess in the outer regions with respect to our sample. This suggests that most of the central mass was already assembled by $z=1$ and the growth in this scenario happened in the outer regions. These outer regions have experienced a growth of 0.4 dex from $0.2\times10^{10}$ M$_\odot$ at $z=1$ to $0.4\times10^{10}$ M$_\odot$ at $z=0$, measured between 8 kpc and R$_{edge}$.
Several studies have shown that the growth of MW-like galaxies at $z < 1$ occurs mostly in the disk \citep[e.g.,][]{Nelson2012, vanDokkum2013, Patel2013, Tan2024a}. For instance, \citet{vanDokkum2013} showed that the stellar mass of galaxies with masses comparable to the MW within 2 kpc remains relatively unchanged since $z=1$ (a change of $0.09\pm0.04$ dex). Meanwhile, the outer regions ($>2$ kpc) grow a factor of $\sim0.25$ dex (see their fig. 4) over this time period. In our case, the central stellar mass growth is $0.076\pm0.001$ dex (within $2$ kpc) and $0.30\pm0.01$ dex (from $2$ kpc to R$_{edge}$), which is consistent with the result of \citet{vanDokkum2013}. This evolution further supports the idea that inside-out growth is the dominant evolutionary mechanism in this scenario \citep[e.g.,][]{Naab2009, Bezanson2009, vanDokkum2010, Trujillo2013}. The pure migration of stars from regions close to the edge of the galaxy at $z=1$ to regions beyond the border is a very unlikely explanation for the substantial mass growth of the outer parts of these galaxies. Therefore, a combination of minor mergers and the transformation of outer galactic gas into new stars is a more likely explanation for the substantial changes we see in the outer regions of disk galaxies from $z=1$ to the present.

Finally, in the right-hand panel of Fig. \ref{stellar-mass-z0}, we make a comparison with $z=0$ galaxies that are 2 to 3 times more massive, consistent with the stellar mass evolution from $z=1$ to $z=0$ reported in previous studies \citep[e.g.,][]{Patel2013, Papovich2015,Tan2024a,Tan2024,Zasowski2025}. As can be seen in the bottom right panel, the ratio of stellar mass density profiles differs at all radii, indicating that the increase in mass occurred throughout the galaxy. However, the difference is more pronounced beyond 8 kpc. The total stellar mass within 8 kpc increases by 0.25 dex at $z = 0$, whereas beyond 8 kpc it grows by up to 0.65 dex. 
\citet{Tan2024a} find that the stellar mass of their MW analogues increases at the same rate in both the inner ( < 2 kpc) and outer ( > 2 kpc) regions. In our case, however, we observe significantly higher growth in the outer regions than in the inner regions. To explain both the inner and outer growth, we need a combination of star formation (inner and outer parts) and galaxy mergers (mainly outer parts). The $z=0$ galaxies have stellar masses of $\sim$10$^{11}$ M$_{\odot}$, and for these massive objects, studies agree in the need for minor merging to progressively create a stellar halo that can explain the evolution of their galaxy outskirts \citep[e.g.,][]{Buitrago2013, vanDokkum2013}.

\samane{Regarding the size evolution, we compare our R$_{edge}$ measurements with those of \cite{chamba2022} for the local galaxies shown in the three mass bins: fixed stellar mass scenario, moderate mass growth, and scenario where they are 2 to 3 times more massive. At  $z \sim 1$, the average R$_{edge}$ of our samples is $\sim$10.5 kpc, while the corresponding medians at $z = 0$ are 15.1 kpc (fixed mass), 21.8 kpc (moderate growth), and 25.9 kpc (mass doubling). These values correspond to size increases of approximately 45$\%$ (0.16 dex) at fixed stellar mass, a factor of $\sim$ 2 (0.31 dex) under moderate mass growth, and up to $\sim$ 2.5 (0.39 dex) in the last scenario. \\
We also computed the effective radius of our galaxies, obtaining R$_e$ = 5.18 kpc for UDF 3372 and R$_e$ = 5.28 kpc for UDF 5417. These values are consistent with the results of \cite{vanderwel2014}, who found typical effective radii of $\sim$5–6 kpc at $z = 1$ for galaxies with stellar masses of $10.5 < \log M_{\star} < 11$. \\ Comparing these two galaxies with those at $z = 0$, we found that, unlike the strong evolution seen in R$_{edge}$, the effective radius changes only slightly from $z = 1$ to $z = 0$. On average, the size changes by approximately 13$\%$, depending on the growth scenario. This shows that the central, half-light region evolves much less than the outer disk, so most of the size growth in MW-like disks since z$\sim$1 happens in their outskirts. Consistent with this picture, we find that the inferred size growth follows the trend shown in Fig. 7 of \cite{buitrago2024}, where the increase in R$_{edge}$ is larger than that measured using R$_e$. Small changes in the effective radius occur because it is, by definition, linked to light concentration. Several studies based on the effective radius \citep[e.g.,][]{Nedkova2021,Kawinwanichakij2021} or R$_1$ \citep{Arjona-Galvez2025} have also found only weak size evolution for disk-dominated galaxies since z $\sim$ 1.}

\section{Summary and Conclusions} \label{sec:summ}
In this study, we analyze two MW–like galaxies with stellar masses of $\sim 4\times10^{10} M_{\odot}$ at $z=1$, taking advantage of the depth and multiwavelength coverage of the JWST and HST datasets. Our goal is to characterize their stellar populations, both within and beyond the galaxy's edge radius, compare them with their $z=0$ counterparts, and gain insight into their evolution over time. To this end, we apply a novel deconvolution method that combines traditional galaxy modeling with advanced PSF deconvolution algorithms. This approach allows us to minimize the scattering of light into the outer regions, which would otherwise mimic an extended stellar halo component.

The edge radius of the galaxies (R$_{edge}$) is at 10.4 kpc for UDF 3372 and 10.6 kpc for UDF 5417, comparable with previous estimates for $z=1$ galaxies \citep{buitrago2024}. The corresponding stellar mass surface densities at the edge ($\Sigma_{edge}$) are $7.6\pm1.3$ and $11.2\pm1.5$ $M_{\odot}/pc^2$ for UDF 5417 and UDF 3372, respectively, in line with \citet{buitrago2024}. Thanks to the depth and multi-wavelength coverage and spatial resolution of JWST and HST, we were able to derive age and metallicity radial profiles extending out to $\sim$15 kpc. Both galaxies show negative metallicity gradients with radius, while the age profiles decline out to $\sim$9 kpc, after which they either turn upward, producing a U-shaped profile (UDF 3372), or remain roughly constant (UDF 5417). 

The stellar population profiles reveal that the minimum stellar age occurs within the edge radius. This suggests that the edge of the galaxy is linked to a change in stellar age.  The continuous decrease in metallicity in the outer regions makes it unlikely that these galaxies grew through the accretion of minor satellites. Therefore, our results imply that the outer disk regions are predominantly populated by older stars that have migrated outwards from the inner disk, as previous studies have suggested \citep[e.g.,][]{roskar2008b, bakos2008color, Azzollini2008, martinz-lombilla2019}. 

By comparing the stellar mass density profiles of our galaxies with those of galaxies at $z=0$ that are about 40\% more massive, we find that the central regions had already stopped growing by $z=1$, whereas the outer regions continued to grow by a factor of two between $z=1$ and $z=0$. By contrast, galaxies that are 2–3 times more massive at $z=0$ show growth in both their inner and outer regions, indicating that such systems continue to accrete material, likely through galaxy interactions or inflows from the intergalactic medium. In both scenarios, it is evident that galaxies have undergone significant structural evolution in their outer disks since $z=1$.

\samane{Our results indicate that while the R$_{edge}$ of MW-like galaxies grows significantly from $z\sim1$ to the present day, the effective radius evolves only weakly. These findings are consistent with an inside-out disk growth scenario, suggesting that most of the structural growth of these galaxies occurs in their outskirts.}

\begin{acknowledgements}
We thank the anonymous referee for carefully reading the manuscript and providing suggestions that helped improve its clarity and strength. SR and FB acknowledge the support of the grants PID2023-150393NB-I00 and CNS2024-154572 from the Spanish Ministry of Science, Innovation, and Universities. Financial support of the Department of Education, Junta de Castilla y Le\'{o}n, and FEDER Funds is gratefully acknowledged (Reference: CLU-2023-1-05). \samane{SR acknowledges financial support from Banco Santander through the Santander Scholarships program (Financial Aid for Predoctoral Research Staff 2025), awarded via the University of Valladolid.} MM acknowledges support from grant RYC2022-036949-I financed by the MICIU/AEI/10.13039/501100011033 and by ESF+, grant CNS2024-154592 financed by  MICIU/AEI/10.13039/501100011033, and program Unidad de Excelencia Mar\'{i}a de Maeztu CEX2020-001058-M. IT acknowledges support from the State Research Agency (AEI-MCINN) of the Spanish Ministry of Science and Innovation under the grant PID2022-140869NB-I00. This research also acknowledge support from the European Union through the following grants: "UNDARK" and "Excellence in Galaxies - Twinning the IAC" of the EU Horizon Europe Widening Actions programs (project numbers 101159929 and 101158446). Funding for this work/research was provided by the European Union (MSCA EDUCADO, GA 101119830). \samane{AAR acknowledges funding from the Agencia Estatal de Investigación del Ministerio de Ciencia, Innovación y Universidades (MCIU/AEI) under grant “Polarimetric Inference of Magnetic Fields” and the European Regional Development Fund (ERDF) with reference PID2022-136563NB-I00/10.13039/501100011033.} 
\end{acknowledgements}

\bibliographystyle{aa} 

\newpage
\appendix
\section{Surface brightness radial profiles for UDF 3372 and UDF 5417} \label{app:sb_prof}

\begin{figure}[h]
    \centering
    \subfigure[UDF 3372]{
        \includegraphics[width=1\linewidth, height=1\linewidth]{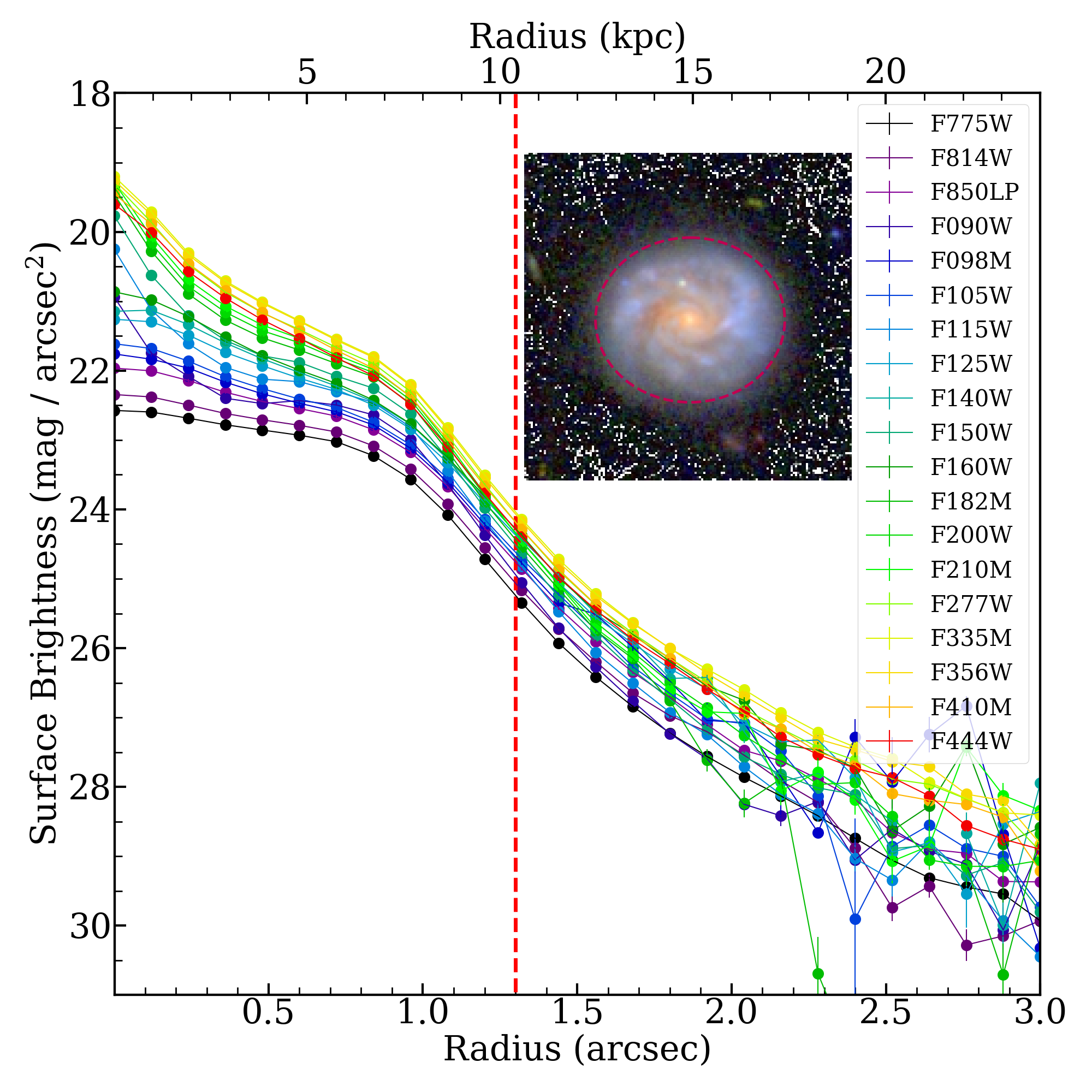}
   }
    \subfigure[UDF 5417]{
        \includegraphics[width=0.5\textwidth]{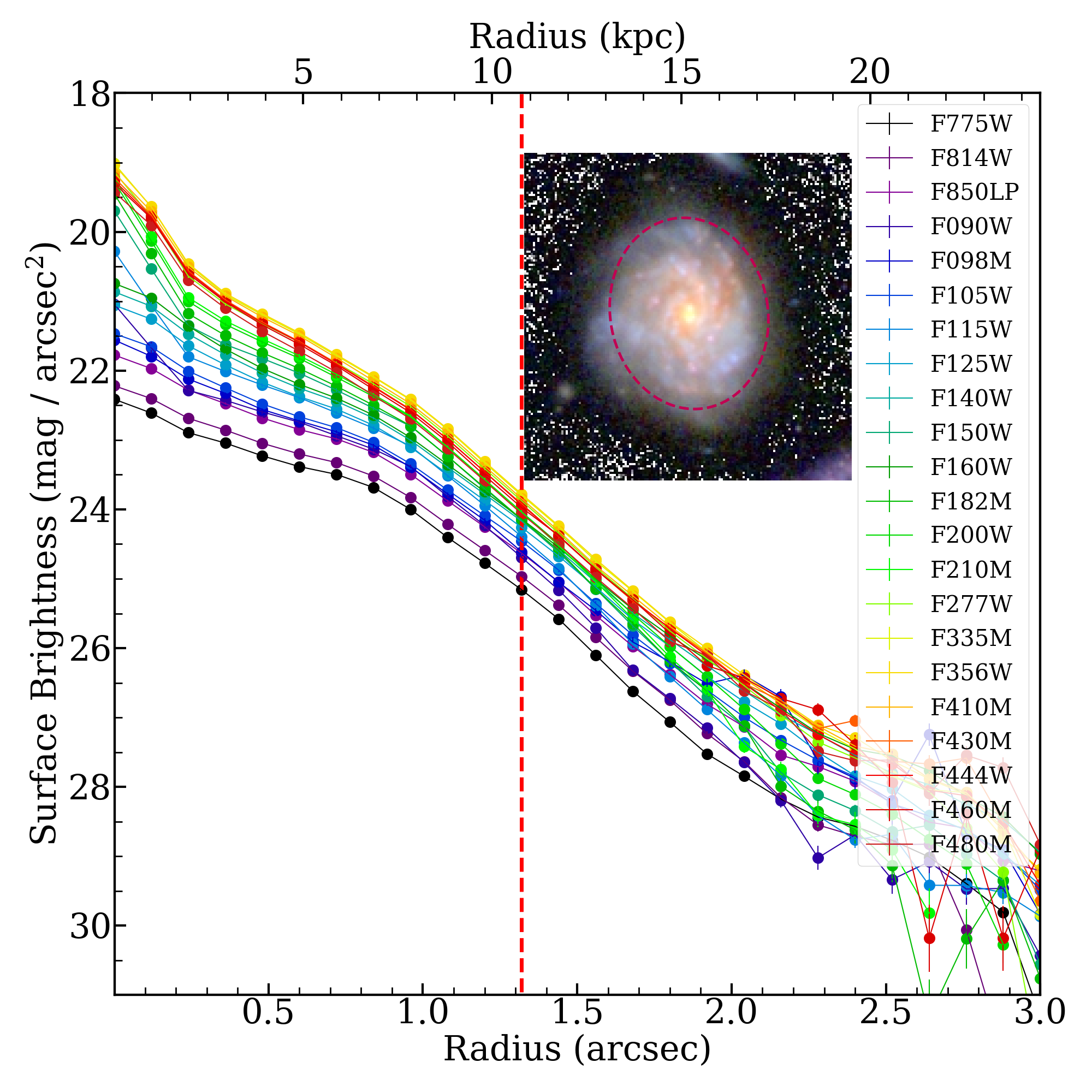}
    }
     \caption{The surface brightness radial profiles of galaxies in all bands are presented. Profiles have been corrected for Galactic extinction and inclination. The dashed red line corresponds to the location of R$_{edge}$. Additionally, a color image of the galaxy is included, with an elliptical aperture that highlights the position of R$_{edge}$. The colour images are constructed using the F090W, F115W, and F150W bands. }
     \label{sb-prof}
\end{figure}

In Fig.~\ref{sb-prof}, we present the surface brightness profiles corrected for Galactic extinction and disc inclination.

\section{Deconvolved surface brightness radial profiles for UDF3372 and UDF5417}
\label{app:dec_prof}
\begin{figure}[h]
    \centering
    \subfigure[UDF 3372]{ \includegraphics[width=1\linewidth,height=1\linewidth]{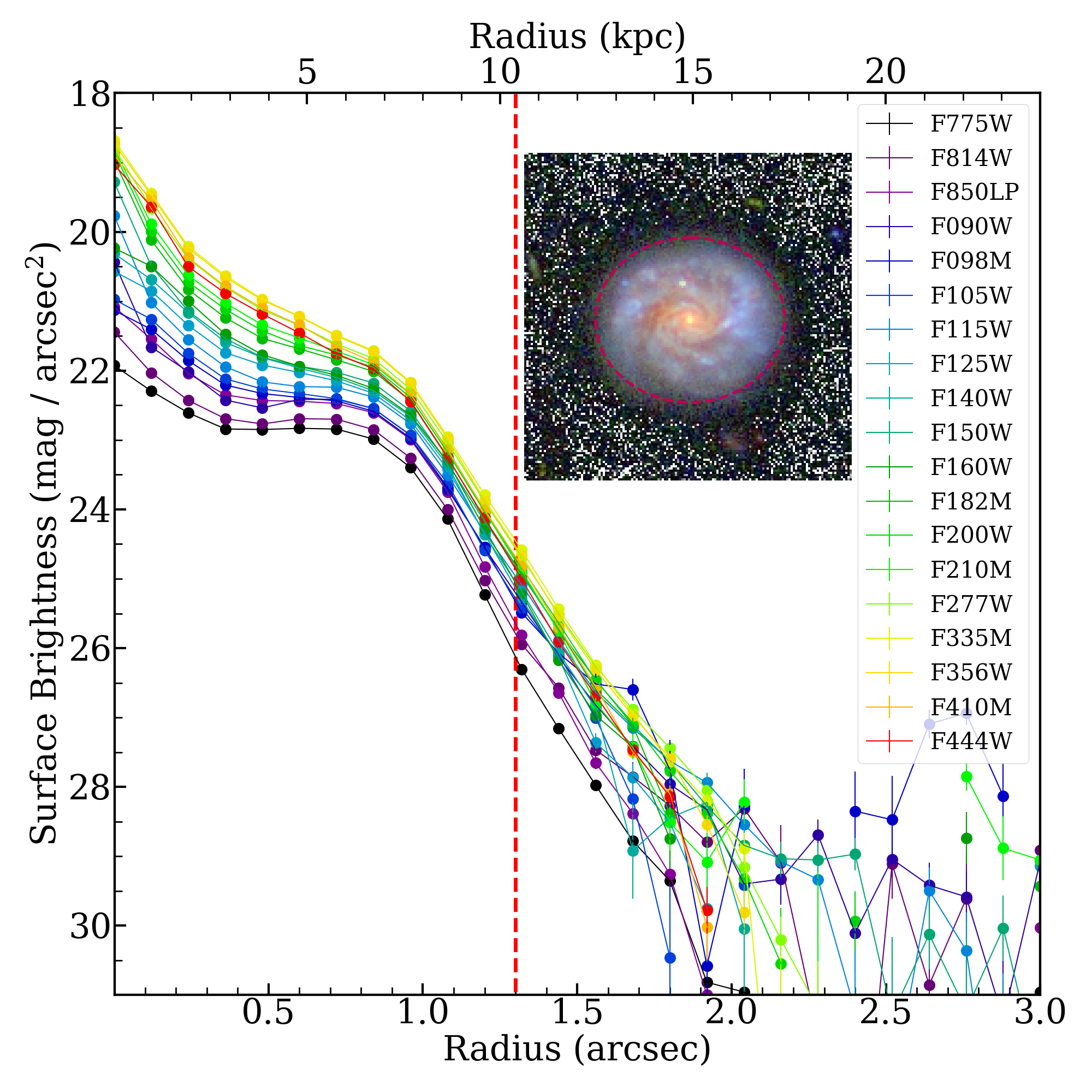}}
    \subfigure[UDF 5417]{
        \includegraphics[width=0.5\textwidth]{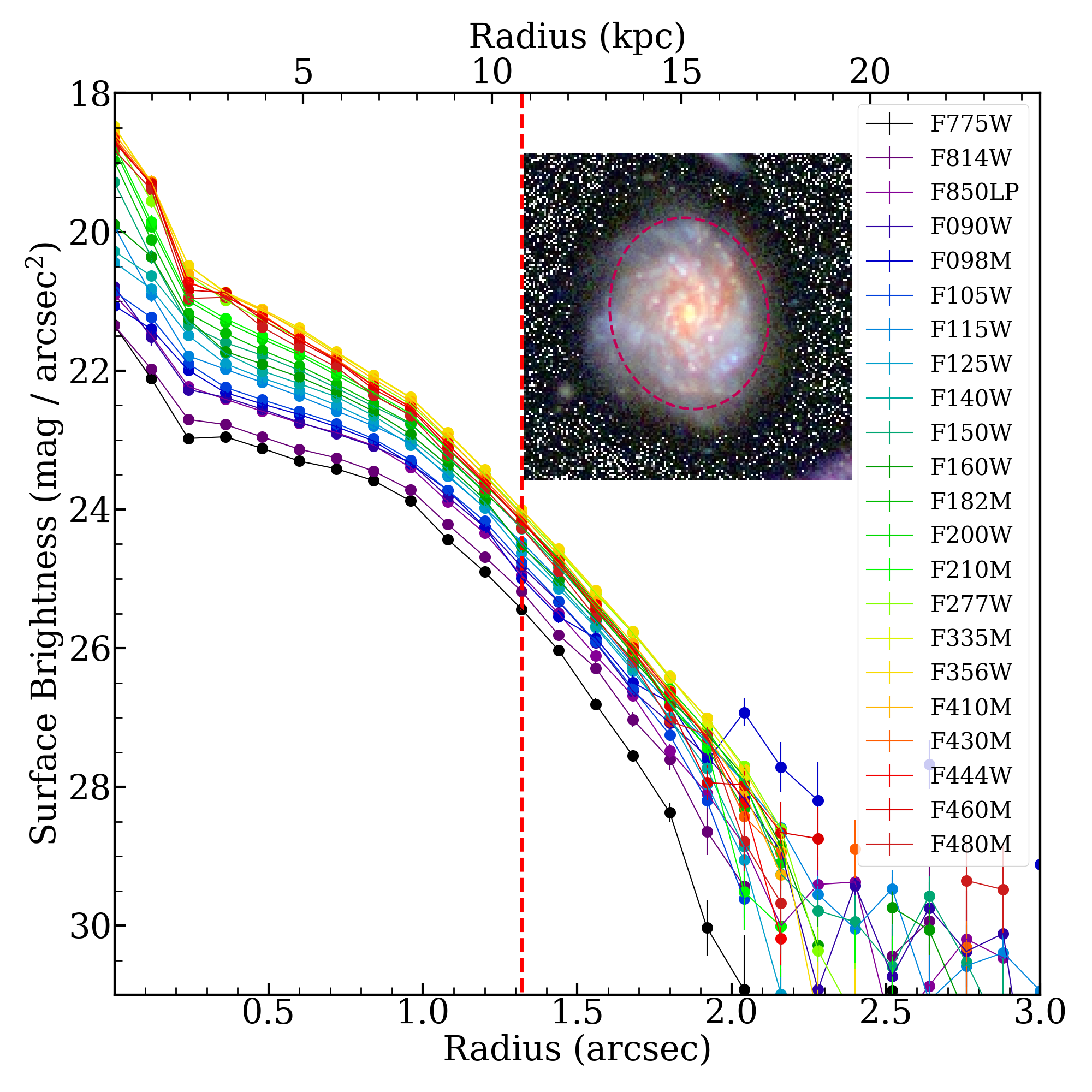}
    }
     \caption{The PSF deconvolved surface brightness radial profiles of galaxies in all bands are presented. The dashed red line corresponds to the location of R$_{edge}$. Additionally, a color image of the galaxy after deconvolution is included, with an elliptical aperture that highlights the position of R$_{edge}$. The colour images are constructed using the F090W, F115W, and F150W bands.}
     \label{decon-prof}
\end{figure}

Fig.~\ref{decon-prof} shows the surface brightness profiles after deconvolving by the effect of the PSF.

\section{ Results using the Wiener-Hunt filter for PSF deconvolution}\label{app:wiener}
The Wiener-Hunt filter is a popular deconvolution technique used to restore images by reducing the effects of blurring and noise. While it may introduce image smoothing, pixel correlation, and changes in background statistics, especially in regions with low surface brightness, it remains robust against issues such as saturation, inaccurate point spread function (PSF) models, and variable background noise. It estimates the original signal by minimizing the mean square error between the true and observed data. In Python, this method can be implemented using functions such as skimage.

\begin{figure}[h]
    \centering
    \subfigure[UDF 3372]{
        \includegraphics[width=0.5\textwidth]{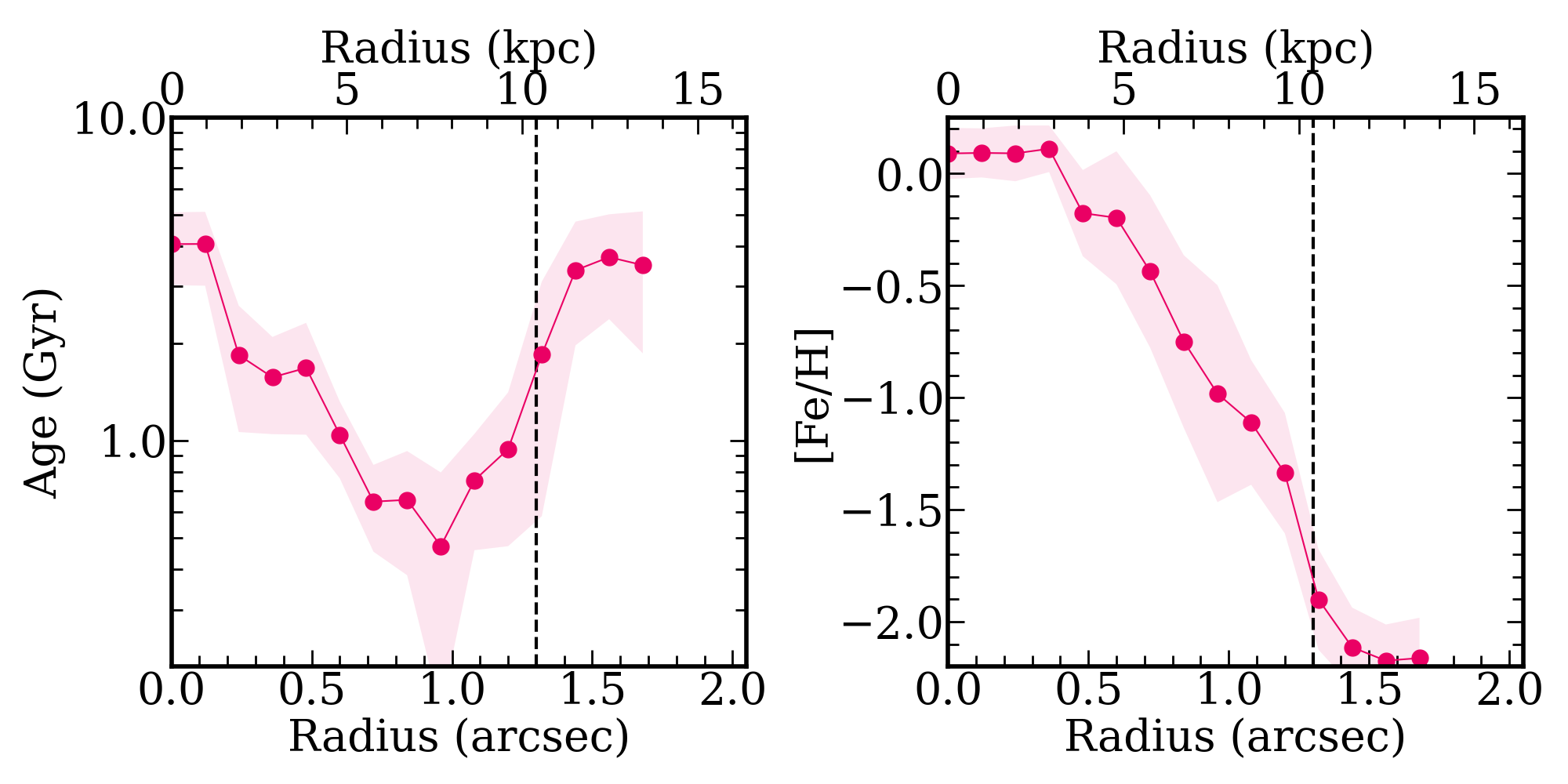}
    }    
    \subfigure[UDF 5417]{
        \includegraphics[width=0.5\textwidth]{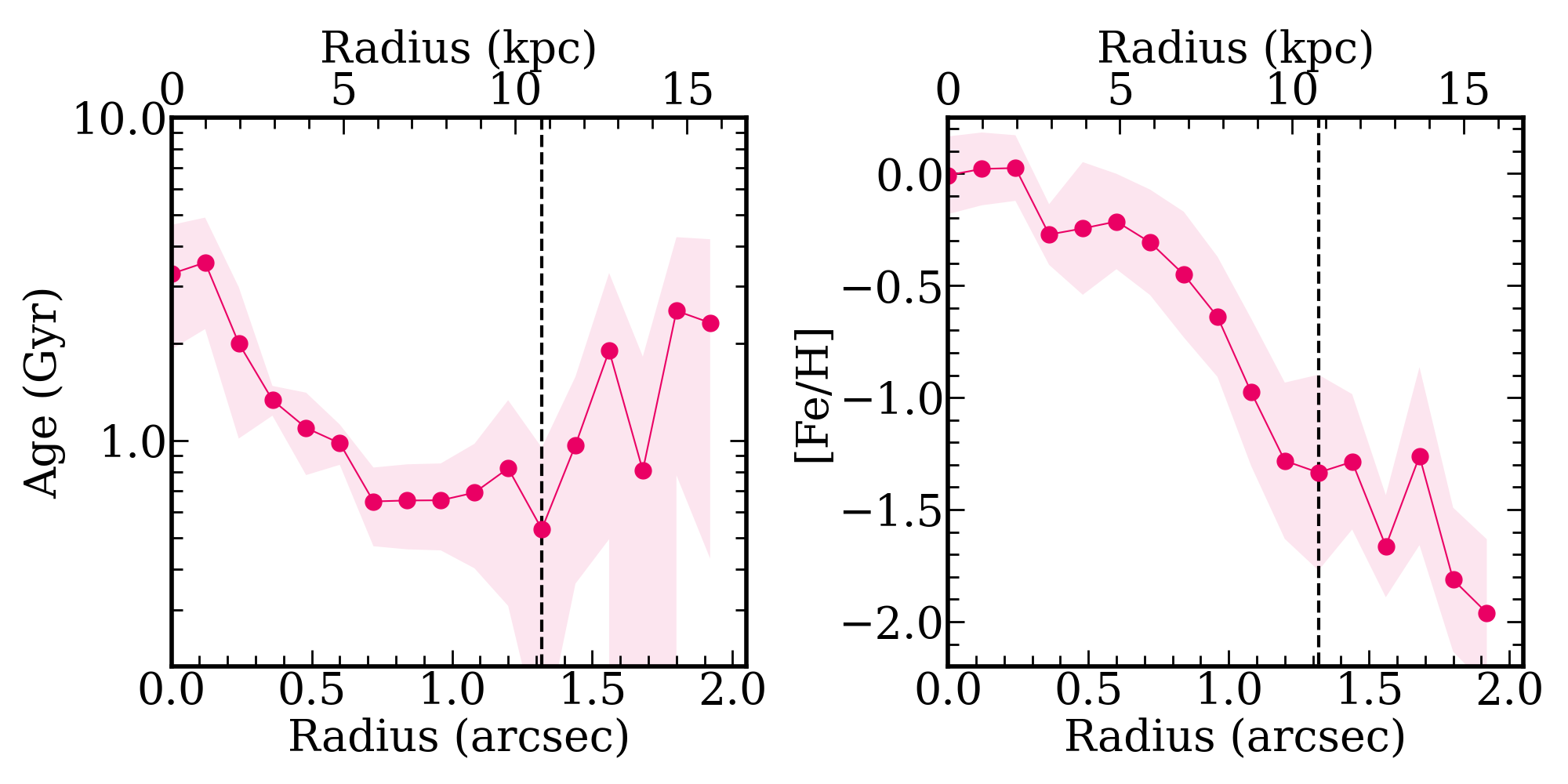}
    }
    \caption{Radial age and metallicity profiles obtained by SED fitting of deconvolved profiles using the Wiener deconvolution algorithm.}
     \label{age-met-wiener}
\end{figure}

In order to evaluate the robustness of our results, we replaced the Wavelet deconvolution with the Wiener deconvolution algorithm. The rest of the methodology used to produce the final deconvolved images was identical to that described in Sec.~\ref{sec:methodology}. We derived age and metallicity profiles from the Wiener-deconvolved images using SED fitting. These profiles (Fig.~\ref{age-met-wiener}) are very similar to those obtained with the Wavelet algorithm (Fig.~\ref{age-met}), indicating the robustness of the results against the choice of deconvolution technique.

\section{Age and metallicity profiles before PSF deconvolution}

\begin{figure}[h]
    \centering
    \subfigure[UDF 3372]{
        \includegraphics[width=0.5\textwidth]{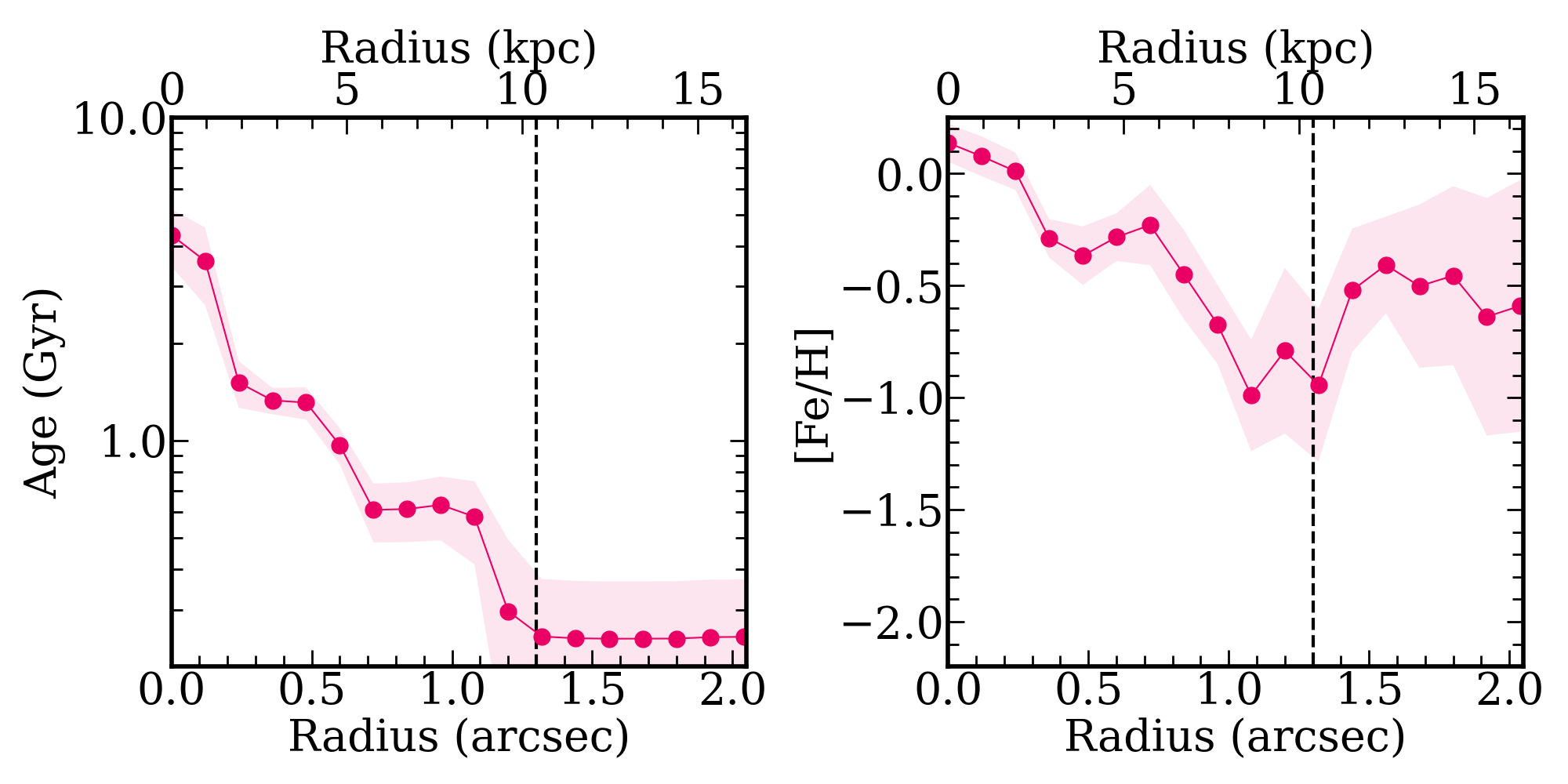}
    }    
    \subfigure[UDF 5417]{
        \includegraphics[width=0.5\textwidth]{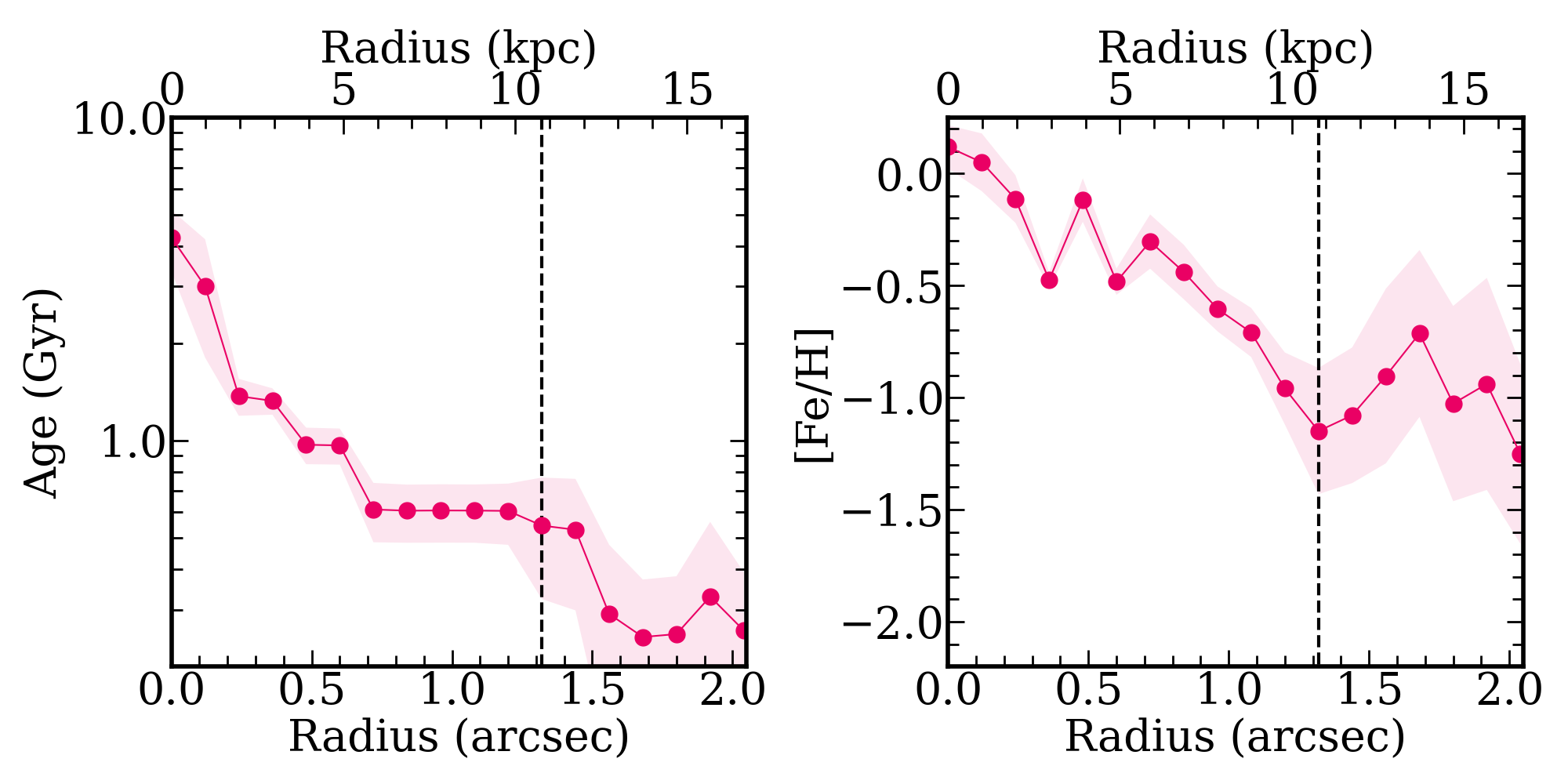}
    }
    \caption{Radial age and metallicity profiles derived from SED fitting of profiles corrected by Galactic extinction, before PSF deconvolution.}
     \label{age-met-PSFin}
\end{figure}

To evaluate how the PSF affects the properties of the stellar population, we estimated the age and metallicity profiles for our galaxies using the surface brightness profiles corrected by Galactic extinction. As shown in Fig.~\ref{age-met-PSFin}, the resulting age profiles indicate older stellar populations in the central regions that become progressively younger with increasing radius. On the other hand, the metallicity profiles show a rise at large radii, indicating high metallicities in the outskirts. This behavior confirms that the PSF significantly affects the observed profiles, particularly in the outer regions, and can mislead the interpretation of stellar population gradients. These results emphasize the importance of applying PSF deconvolution before SED fitting to obtain reliable, physically meaningful age and metallicity profiles. 

\section{Weighted mean profile contributions}\label{average profiles}
To build the average profiles at each radial distance in Fig. \ref{stellar-mass-z0}, the uncertainties of the individual profiles of \cite{chamba2022} were taken into account. Those with smaller error bars are given more weight than those with larger ones. Thus, the average profiles are the weighted mean of the individual galaxies. Specifically, we have applied the following equation:

\begin{equation}
    <\Sigma_*(R)>=\frac{\sum\limits_{i=1}^{n}w_i(R)\Sigma_{*,i}(R)}{\sum\limits_{i=1}^{n}w_i(R)}
\end{equation}

where $w_i(R)=(\sigma_{min}(R)/\sigma_{i}(R))^2$. $\sigma_{min}(R)$ is the lower error bar of the profiles combined at given $R$. The uncertainty on the weighted mean profiles was estimated by combining two contributions. First, the statistical error of the weighted mean at each radial bin was calculated as:

\begin{equation}
    \sigma^2_{ave}(R)=\frac{1}{\sum\limits_{i=1}^{n}1/\sigma^2_{i}(R)}
\end{equation} 

where $\sigma^2_{i}(R)$ represent the uncertainties of the individual profiles. Secondly, we quantified the intrinsic scatter of the stacked profiles at each radius by calculating the standard deviation.

\begin{equation}
\sigma_{\mathrm{int}}(R)
= \mathrm{std}\!\big(\Sigma_{\ast,i}(R)\big)
\end{equation}

The total error bar associated with the weighted mean profile was then calculated as the square root of the sum of the squares of these two terms.

\begin{equation}
\sigma_{\mathrm{tot}}(R)
= \sqrt{\sigma_{\mathrm{ave}}^2(R) + \sigma_{\mathrm{int}}^2(R)}
\end{equation}

This approach ensures that the resulting average profiles are statistically representative, taking into account the different levels of error in the individual measurements.

\section{Stellar masses }
We computed the total stellar mass and the stellar mass enclosed within an 8 kpc radius for galaxies at $z=1$ and in the present-day Universe ($z=0$). The total stellar mass was calculated up to the outermost radius covered by the available data in the profiles. For the $z=0$ galaxies, we first constructed an average stellar mass surface density profile by stacking individual profiles. Then, we integrated this average profile to estimate the total stellar mass and the mass enclosed within 8 kpc. This analysis was performed on two groups of galaxies, as shown in Fig.~\ref{stellar-mass-z0}: Group 1 considers galaxies that are $50\%$ more massive than those at $z = 1$. Group 2 considers galaxies that are twice as massive as those at $z = 1$.

\begin{table}[h]
\centering
\caption{Total stellar masses and mass within 8 kpc.}\label{tab:masses}
\begin{tabular}{|c|c|c|}
\hline
Galaxy  & \makecell{Total stellar mass\\(M$_\odot$)} & \makecell{Stellar mass in\\R$\leq$8 kpc (M$_\odot$)}  \\ \hline
UDF 5417       & $4.1\times10^{10}$  & $3.53\times10^{10}$  \\  \hline
UDF 3372       & $3.9\times10^{10}$  & $3.49\times10^{10}$ \\ \hline
Group 1  & $5.8\times10^{10}$ & $4.2\times10^{10}$ \\\hline
Group 2  & $8.5\times10^{10}$  & $6.1\times10^{10}$ \\ \hline    
\end{tabular}
\tablefoot{We also include the average total stellar mass and within 8 kpc of a sample of galaxies at $z=0$ in two groups, Group 1, those that have $40-50\%$ more mass than $z=1$ and Group 2, for those that are 2.5 to 3 times more massive than $z=1$, shown in Fig. \ref{stellar-mass-z0}.}
\end{table}

\section{Internal dust effect} \label{dust}
\begin{figure}[h!]
    \centering
    \subfigure[Moderate internal dust correction for UDF 3372]{
        \includegraphics[width=0.5\textwidth]{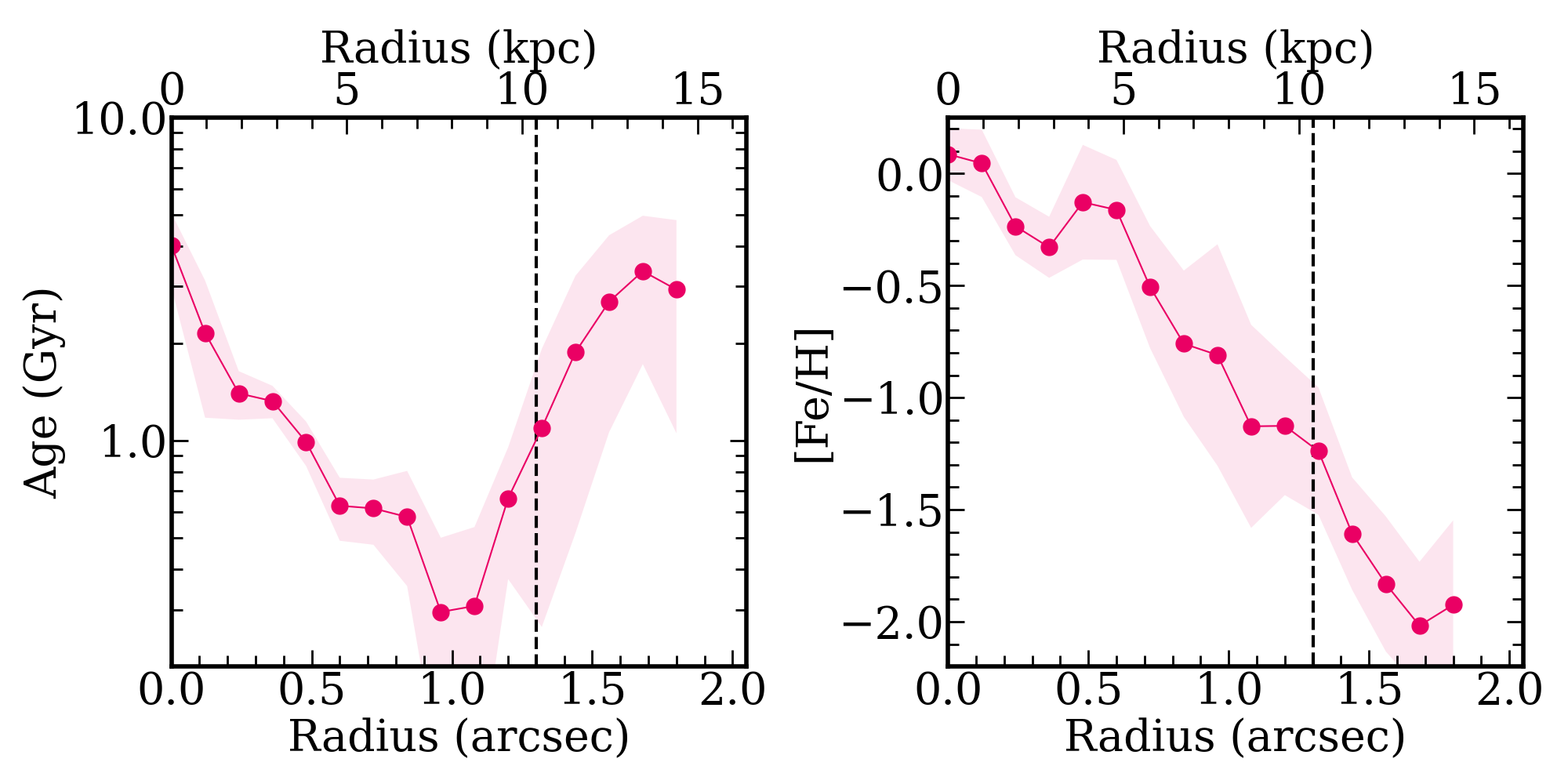}}    
    \subfigure[Strong internal dust correction for UDF 3372]{
        \includegraphics[width=0.5\textwidth]{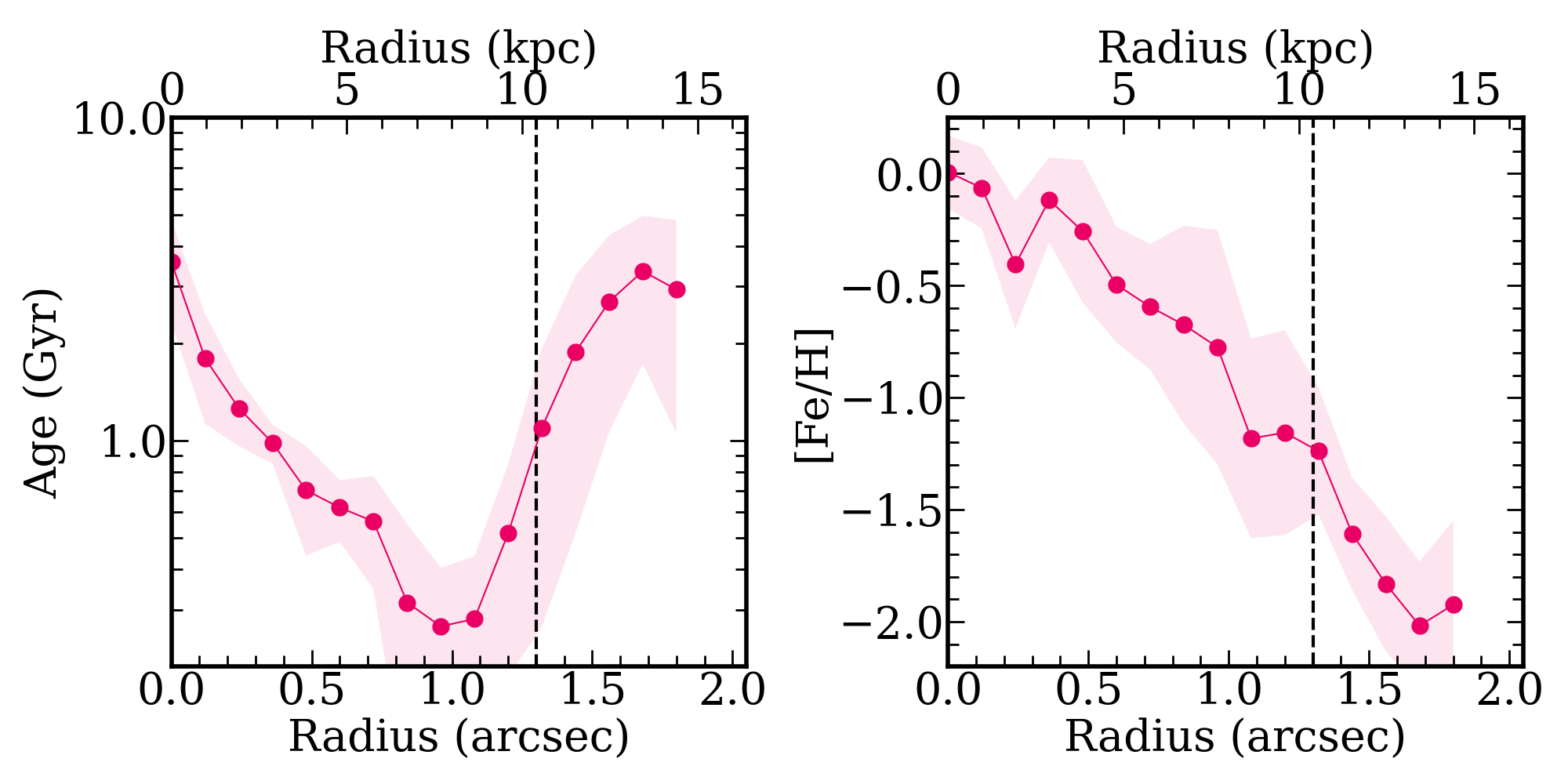}
    }
    \subfigure[Moderate internal dust correction for UDF 5417 ]{
        \includegraphics[width=0.5\textwidth]{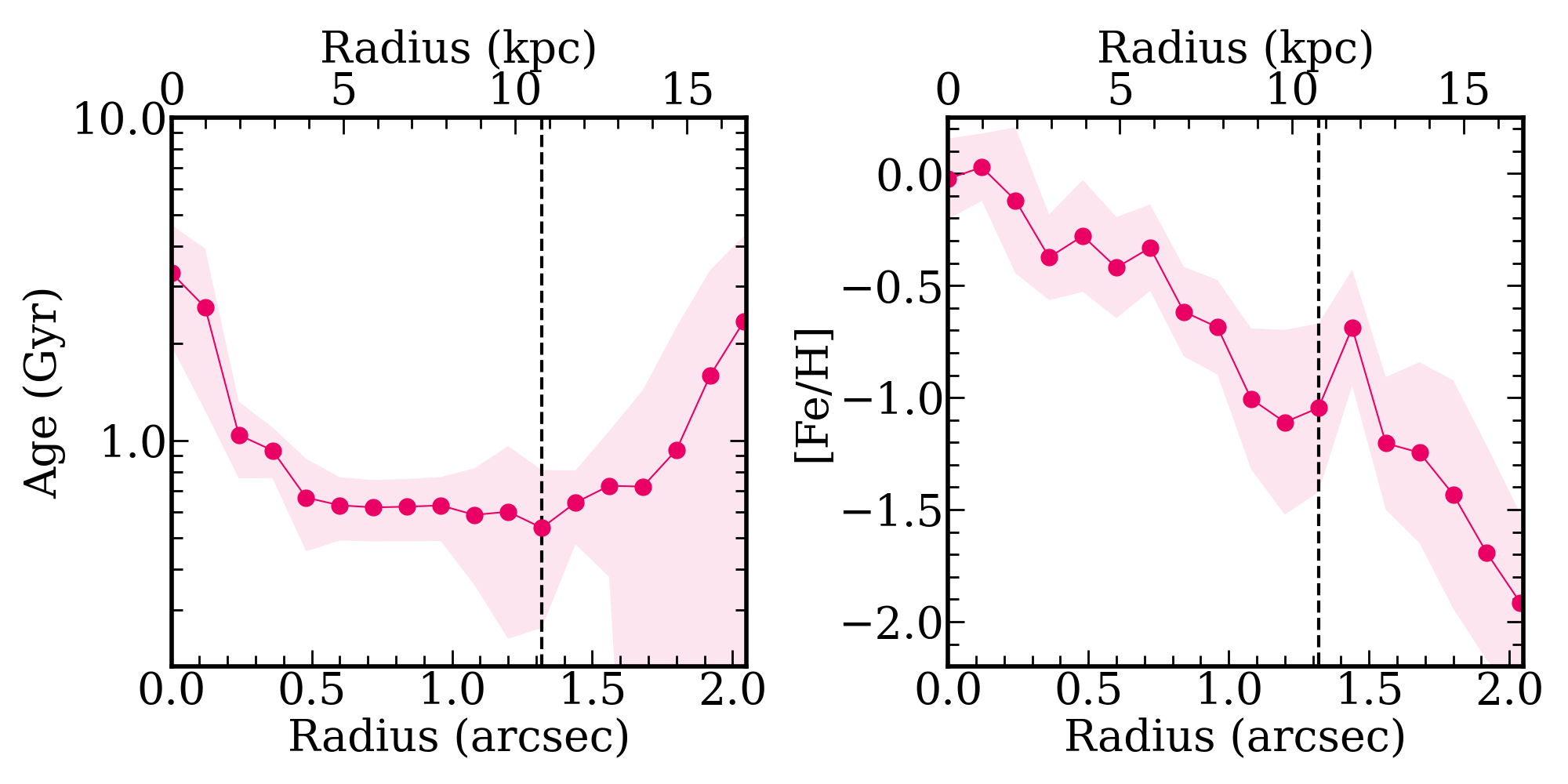}
    }    
    \subfigure[Strong internal dust correction for UDF 5417]{
        \includegraphics[width=0.5\textwidth]{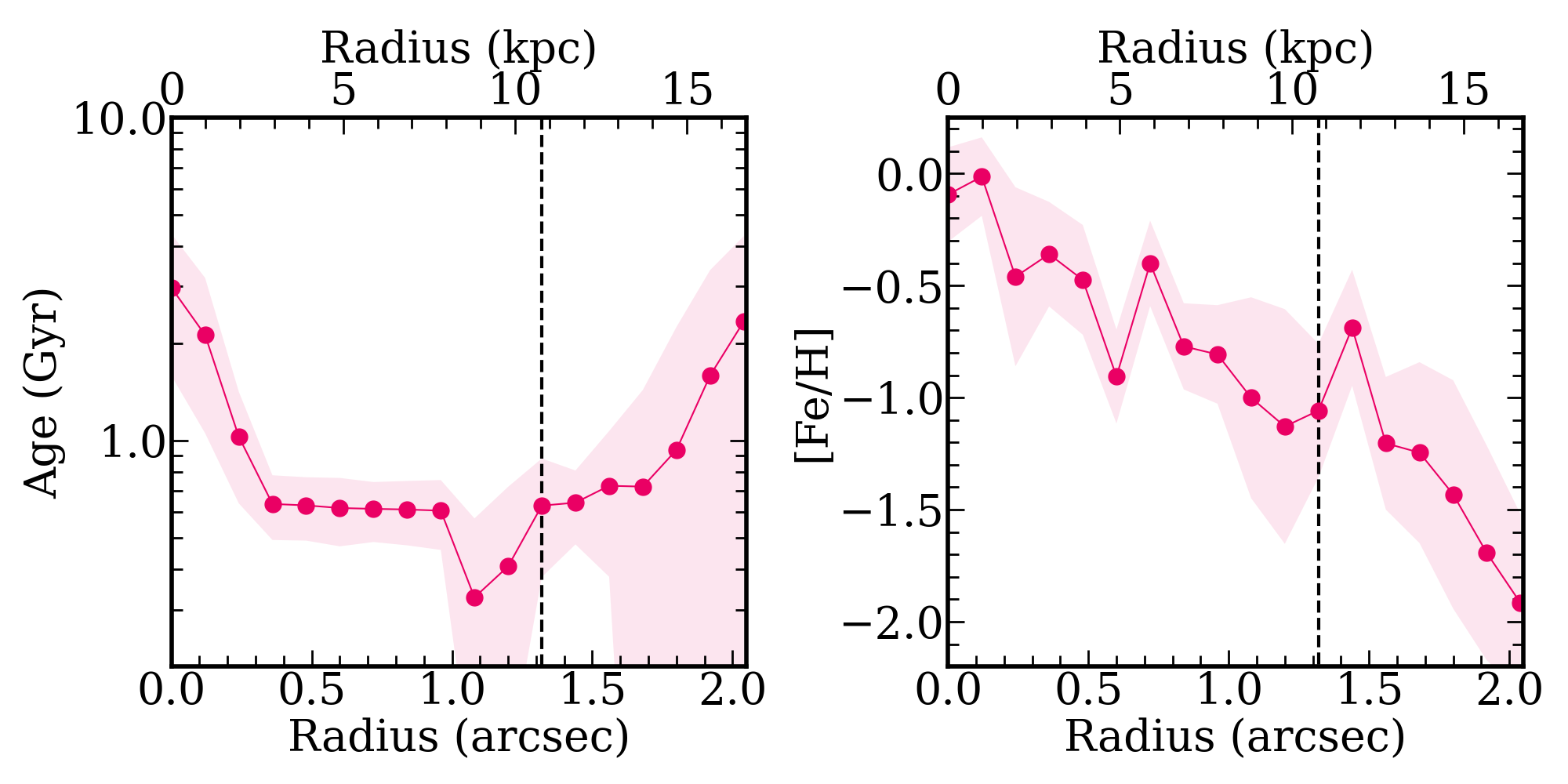}
    }
    \caption{Radial age and metallicity profiles after dust correction for UDF 3372 and UDF 5417.}
     \label{dust-correction-33372}
\end{figure}

 We performed two tests to quantify its potential impact. For simplicity, we used the correction coefficient presented only for the disk, while accounting for different optical depths across different parts of galaxies. This takes into account the expected lower amount of dust in the younger selected regions.  For the first test (moderate dust correction), we assumed a central face-on B-band optical depth($\tau_B^f$) of 4 within the inner 5 kpc, 1 in the outer disk region (5–10 kpc), and 0.1 beyond R$_{edge}$. For the second case (strong dust correction), we used larger optical depths of 8, 2, and 0.1 for the same radial ranges (inner 5 kpc, 5–10 kpc, and beyond R$_{edge}$, respectively). The resulting effects of both correction schemes on the derived age and metallicity are presented in Fig. \ref{dust-correction-33372}.

\begin{figure}[h]
    \centering
    \subfigure[UDF 3372]{
        \includegraphics[width=0.4\textwidth]{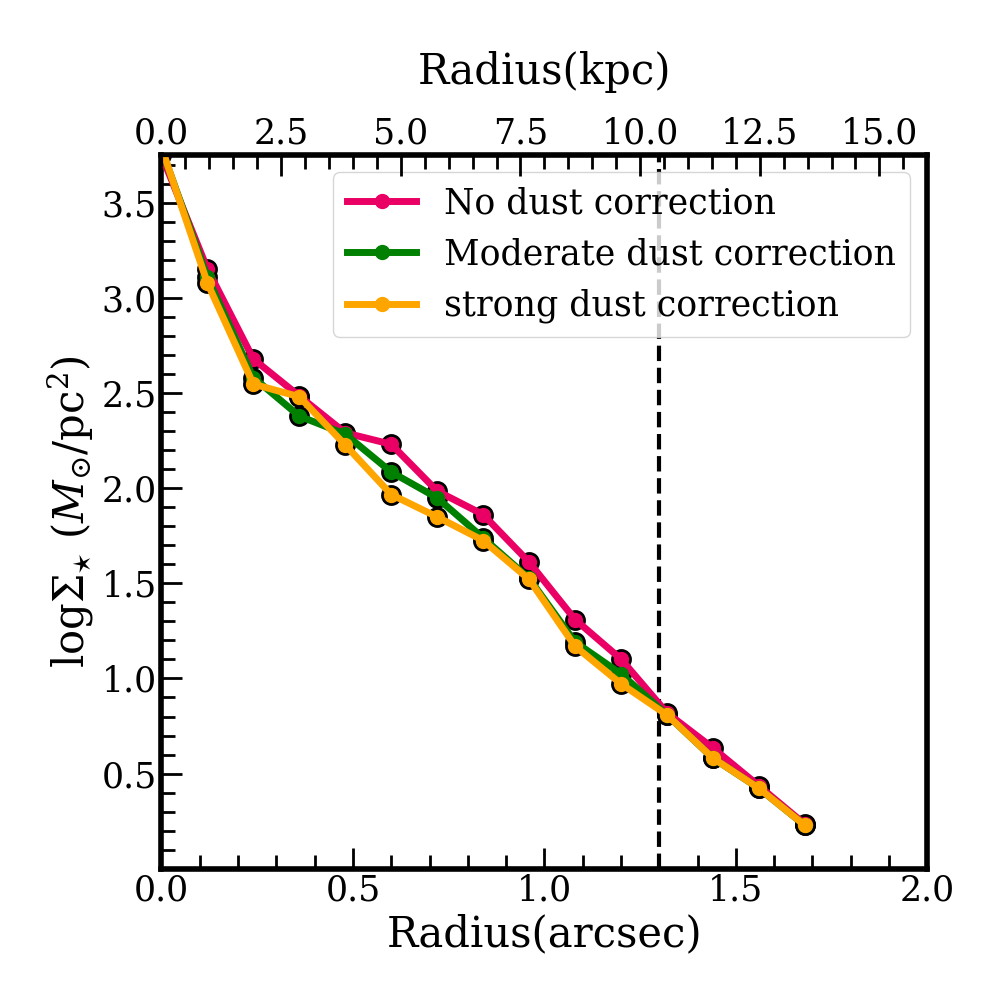}
    }    
    \subfigure[UDF 5417]{
        \includegraphics[width=0.4\textwidth]{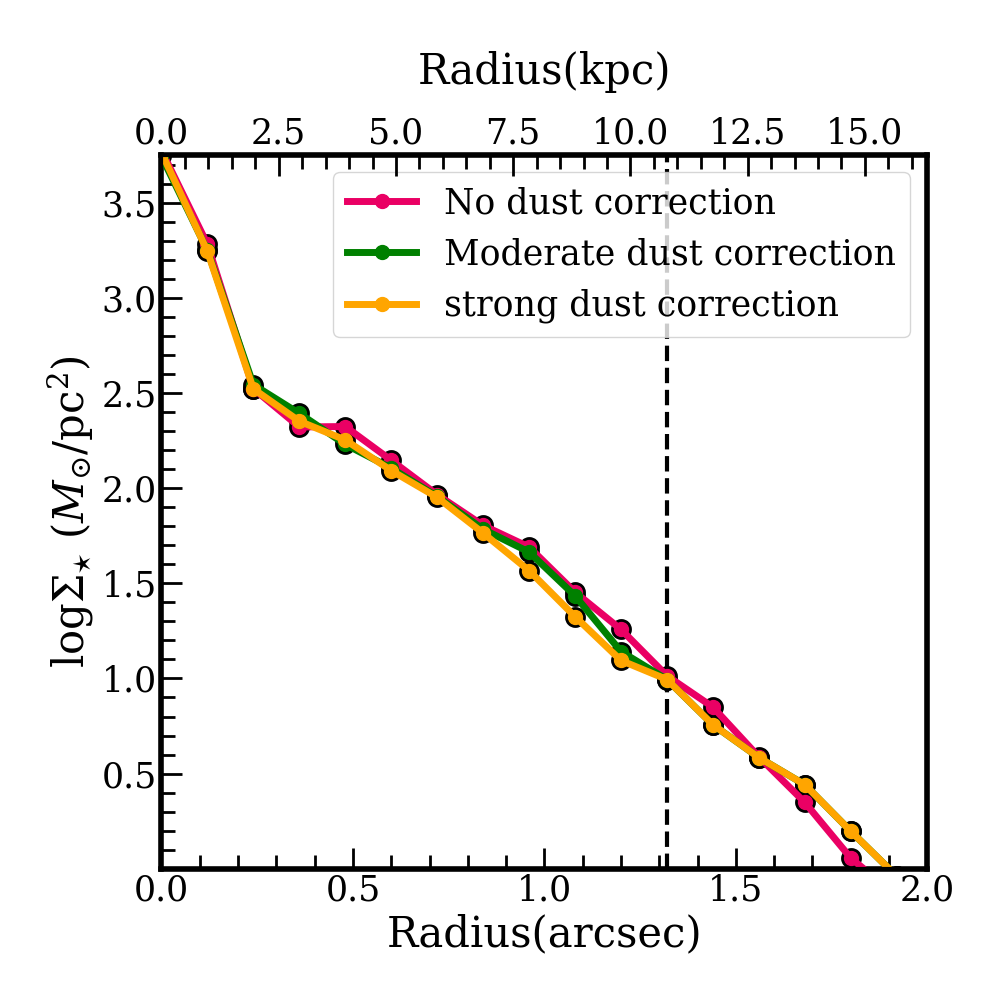}
    }
    \caption{Comparison of the stellar mass profiles obtained without internal dust attenuation correction and after applying moderate and strong corrections.}
     \label{mass-dust-corr}
\end{figure}

We also compared the stellar mass profiles with and without internal dust correction, as shown in Fig. \ref{mass-dust-corr}. Internal dust corrections change the stellar mass of UDF 3372 by only about $0.04-0.06$ dex ($9-13\%$) for moderate and strong corrections, and for UDF 5417 by only $0.01-0.03$ dex ($3-8\%$). This shows that internal dust has a minimal effect and does not change our conclusions.

\begin{figure}[h]
    \centering
    \subfigure[UDF 3372]{
        \includegraphics[width=0.5\textwidth]{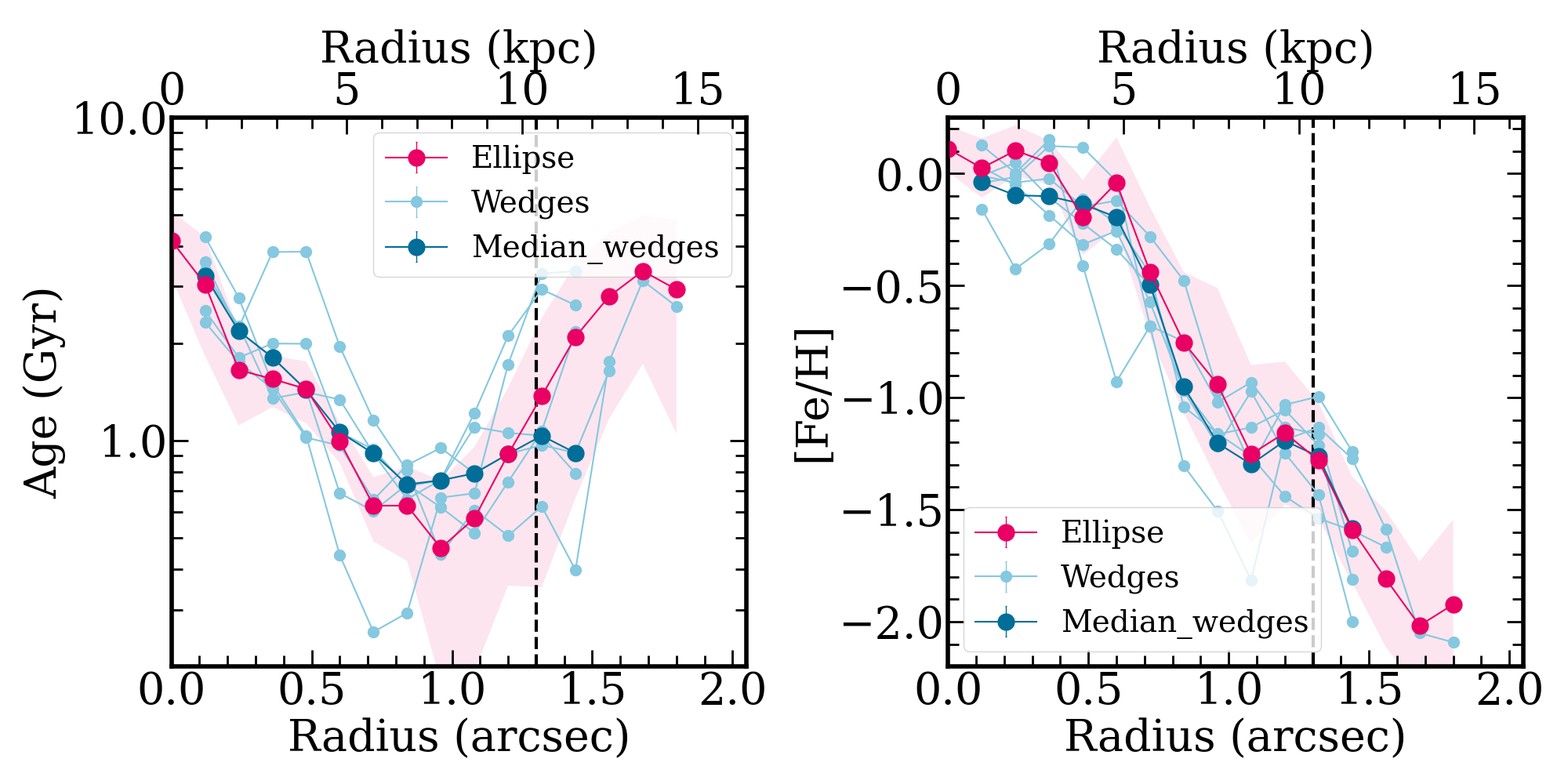}
    }    
    \subfigure[UDF 5417]{
        \includegraphics[width=0.5\textwidth]{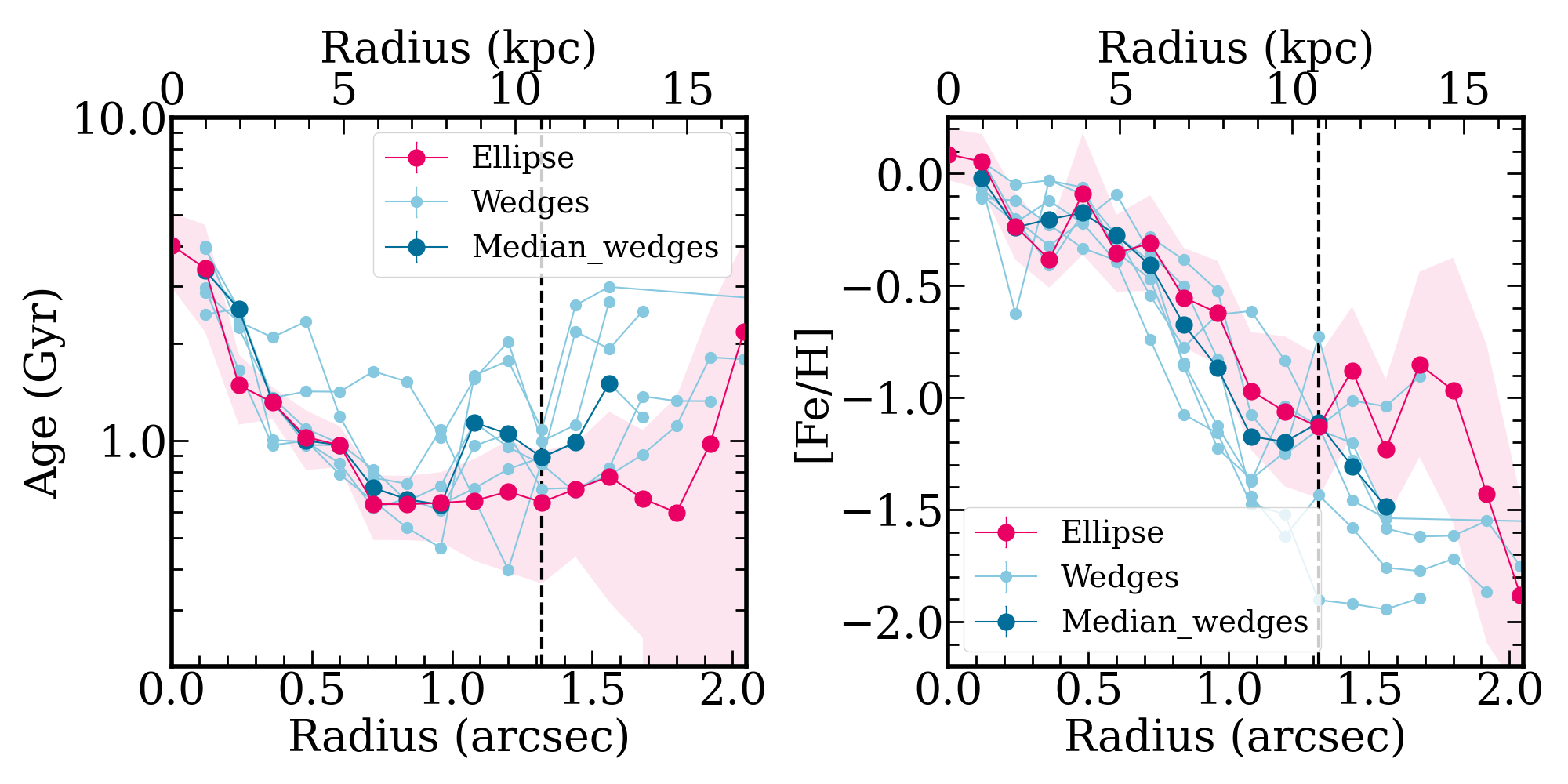}
    }
    \caption{Radial age and metallicity profiles using different angular vectors (wedges).}
     \label{wdge-prof}
\end{figure}

\samane{\section{Wedges age and metallicity profiles}\label{wedge_age_met}
Fig \ref{wdge-prof} shows the comparison between age and metallicity profiles using the ellipse and wedge methods. The wedges are 60 degrees wide. The median of all wedge profiles has also been shown in the plots. }

\section{ Analogous to Fig. \ref{psf_corr} and Fig. \ref{SEDfig} for UDF 3372}

\begin{figure*}[h]
    \centering \includegraphics[width=\linewidth]{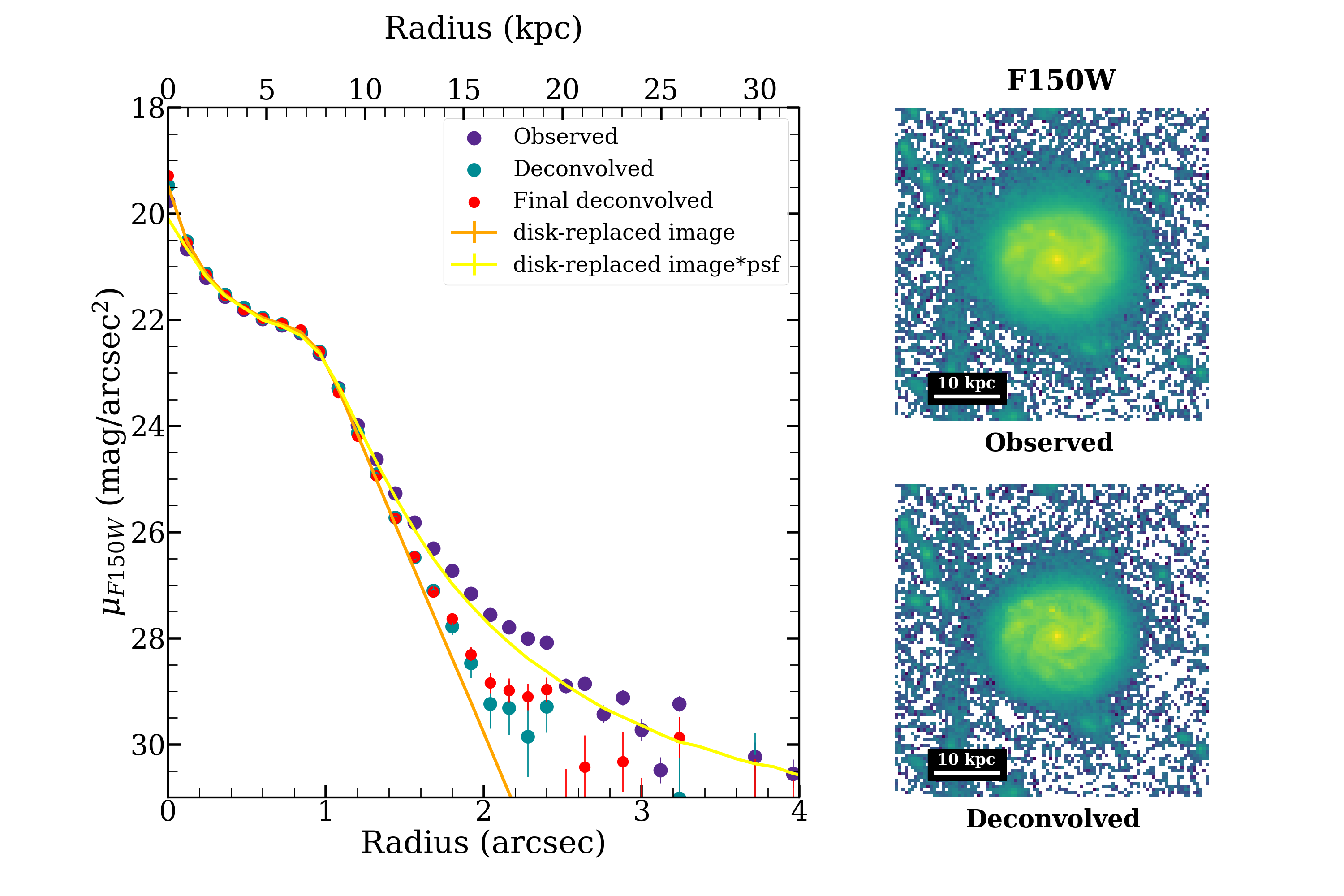}
    \caption{Left panel: Surface brightness profiles of the galaxy UDF 3372 for the F150W band. This figure illustrates the different steps in the deconvolution process. We start with the observed galaxy (purple circles) and deconvolve it using the Wavelet deconvolution (teal circles). We then replace the inner parts of the model with the galaxy deconvolved (disk-replaced image, orange line) and convolve it with the image's PSF (yellow line). The final step is to obtain the residuals (observed - convolved disk-replaced image) to add them back to the model galaxy to obtain the final deconvolved image of the galaxy (red circles).
    Right panels: The galaxy UDF 5417 in the F150W band. The top panel shows the observed image of the galaxy, and the bottom panel shows the final deconvolved image.}
    \label{psf_corr_3372}
\end{figure*}

\begin{figure*}[h]
    \centering \includegraphics[width=1\linewidth,height=1\linewidth]{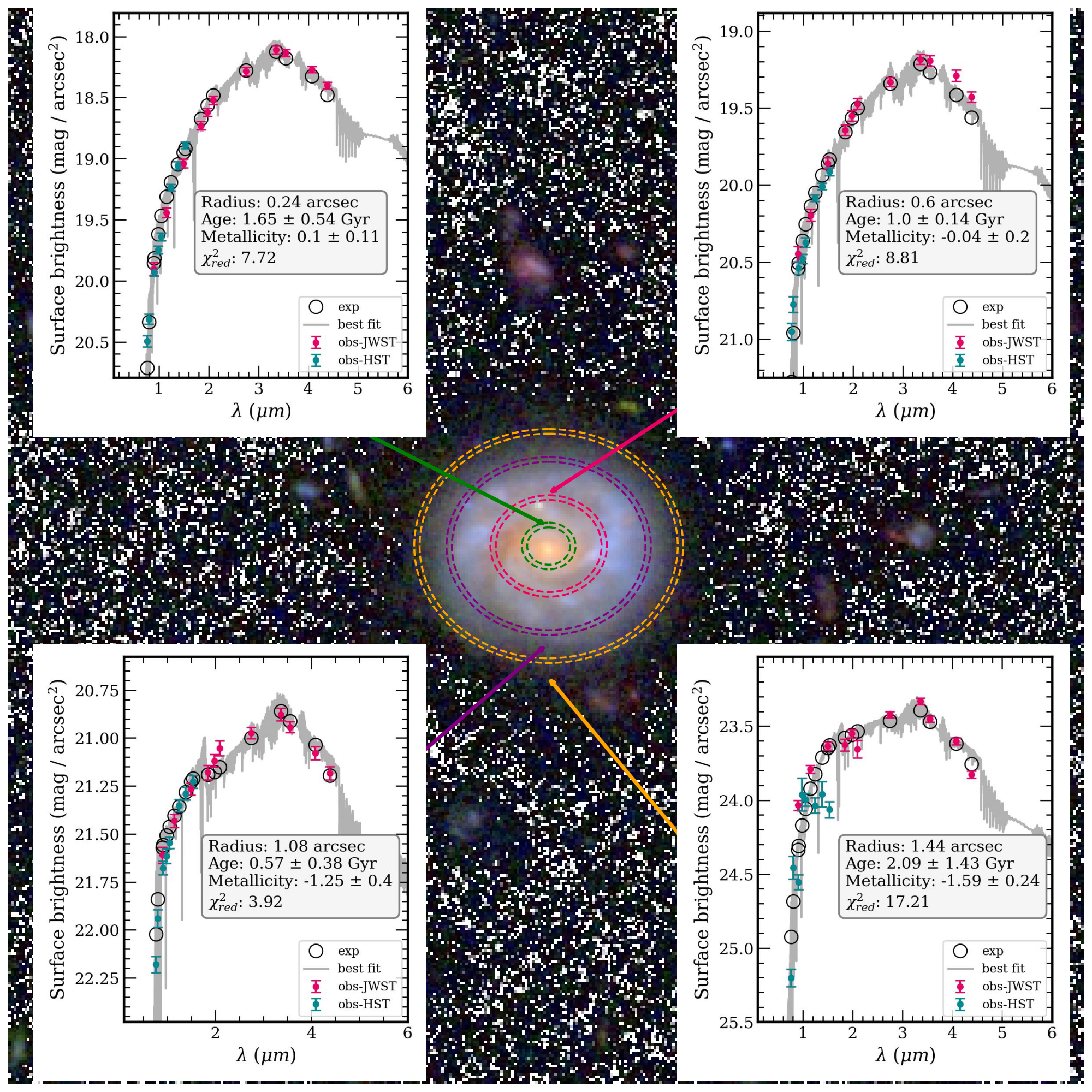}
    \caption{Example of 4 SEDs at different radial distances for the galaxy UDF 3372. The figure shows the SEDs from a region close to the central core (top left) to the outer regions (bottom right), as indicated by the colours of the ellipses. In each panel, the blue and red circles represent the photometry of the galaxies (HST and JWST filters, respectively). The grey continuum is the best-fit SSP model, and the open circles indicate the convolution of the model with the response of the filters used here. The reduced $\chi^2$, estimated age, metallicity, and distance are reported in each plot. }
    \label{SEDfig-3372}
\end{figure*}

\end{document}